\documentclass[aps,prl,twocolumn,showpacs,10pt,floatfix,longbibliography]{revtex4-2}

\usepackage{graphicx}
\usepackage{subfigure}
\usepackage{amsmath}
\usepackage{amsfonts}

\usepackage{mathrsfs}
\usepackage{bbm}
\usepackage{subfigure}
\usepackage{xcolor}
\usepackage[colorlinks=true, urlcolor=blue, linkcolor=blue, citecolor=blue]{hyperref}

\usepackage[T1]{fontenc}

\usepackage{siunitx}
\usepackage{bm}
\usepackage[version=4]{mhchem} 

\usepackage{stackengine}[2013-09-11]
\newcommand\slashzero{\stackinset{c}{}{c}{}{/}{0}}

\usepackage{dsfont}

\AddToHook{cmd/appendix/before}{%
  \setcounter{equation}{0}%
  \setcounter{theorem}{0}%
}

\begin{document}

\title{Transverse Josephson diode effect in tilted Dirac systems}

\author{W.~Zeng}
\email[E-mail: ]{zeng@ujs.edu.cn}
\affiliation{Department of physics, Jiangsu University, Zhenjiang 212013, China}

\begin{abstract}

We theoretically study the transverse charge transport in Josephson junctions based on the tilted Dirac materials with valley-dependent gaps. It is shown that a finite tilt-assisted transverse Josephson Hall current is present under broken time-reversal symmetry. This transverse current is driven by the superconducting phase difference across the junction and exhibits a nonsinusoidal current-phase relation, leading to the transverse Josephson diode effect (TJDE), where the critical currents flowing oppositely along the transverse direction are asymmetric. Compared to the conventional longitudinal Josephson diode effect, the predicted TJDE supports a fully polarized diode efficiency with a $100\%$ quality factor and can completely decouple the input signal path from the output, suggesting potential applications for nonreciprocal superconducting devices.

\end{abstract}
\maketitle

\textit{Introduction.}---The nonreciprocal charge transport in Josephson junctions manifests as direction-selective critical supercurrents $j_{c}^+\neq j_{c}^-$ with $j_{c}^{+(-)}$ being the critical supercurrent along the positive (negative) direction of the junction, which is termed as the Josephson diode effect \cite{PhysRevX.12.041013,PhysRevLett.99.067004,PhysRevB.110.014519,PhysRevLett.130.266003}. The Josephson diode effect has been observed or proposed in many systems that break both inversion symmetry (IS) and time-reversal symmetry (TRS) \cite{PhysRevLett.129.267702,PhysRevLett.131.096001,PhysRevB.109.075412,PhysRevB.108.054522,PhysRevB.109.L081405}. The ability of the Josephson diode to conduct current along one direction is characterized by its quality factor \cite{PhysRevB.109.094518,PhysRevB.110.134511,PhysRevB.106.214524}. Up to now, some highly efficient Josephson diode effects with large quality factors have been reported. The quality factor can be $90\%$ in the Josephson diode effect proposed via asymmetric spin-momentum locking states \cite{PhysRevApplied.21.054057} and can even reach $100\%$ by exploring a simple supercurrent network in the field-free graphene Josephson triode system \cite{chiles2023nonreciprocal}.

However, previous works on the Josephson diode effect focused on the longitudinal direction of the junction, where the output rectified signal is in the direction parallel to the input. A natural question to ask is whether this nonreciprocal diode effect can be generated in the transverse direction of the Josephson junction, so that the paths of the input excitation and the output rectified signal can be decoupled. Recent studies reported that the tilted fermions may induce the transverse transport in tunnel junctions, such as the tunneling valley Hall effect \cite{PhysRevLett.131.246301}, the tunneling chirality Hall effect \cite{PhysRevB.110.024511}, and the nonlinear valley Hall effect \cite{PhysRevLett.132.096302}, suggesting that the tilted Dirac system serves as a promising platform for designing transverse superconducting devices.

Most of the Dirac materials possess the isotropic spectrum in momentum space, where the Dirac cones are symmetric \cite{RevModPhys.92.021003,RevModPhys.90.015001,PhysRevLett.97.067007,Zeng_2022,HUANG2023128671}. However, a finite tilt of Dirac cones can occur since the Lorentz symmetry is not necessarily the symmetry group in condensed matter systems \cite{PhysRevB.100.045144}, such as in the case of strained graphene \cite{PhysRevB.78.045415}, $8$-\textit{Pmmn} borophene \cite{PhysRevLett.112.085502,PhysRevB.93.241405,PhysRevB.97.125424}, organic conductor $\alpha\textnormal{-}\ce{(BEDT\textnormal{-}TTF)2I3}$ \cite{doi:10.1143/JPSJ.75.054705}, and $1T'$ monolayer transition metal dichalcogenides \cite{PhysRevB.103.125425,sxss5}. Moreover, a finite valley-dependent band gap can also be generated in the tilted Dirac systems. The \textit{ab initio} calculation suggests that a band opening in $8$-\textit{Pmmn} borophene can be achieved by hydrogenation and controlled by strain \cite{doi:10.1021/acsaelm.9b00017}, and the massive fermions in $\alpha\textnormal{-}\ce{(BEDT\textnormal{-}TTF)2I3}$ has been experimentally reported \cite{doi:10.1143/JPSJ.75.054705}. In addition, the off-resonant circularly polarized radiation accompanied by a sublattice staggered potential is also an alternative method for producing the valley-dependent mass in the regime of Floquet picture \cite{PhysRevB.103.165415,PhysRevB.102.045417}.

\begin{figure}[tp]
\begin{center}
\includegraphics[clip = true, width =\columnwidth]{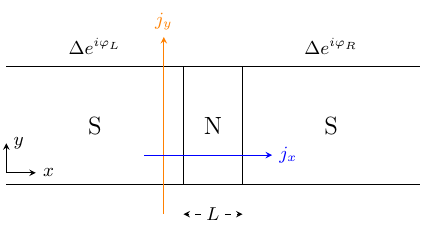}
\end{center}
\caption{Schematic illustration of the Josephson junction based on the tilted Dirac material. The longitudinal direction of the junction is along the $x$ axis. The superconducting regions (denoted by S) are located at $x<0$ and $x>L$, where the $s$-wave pairing potentials are $\Delta e^{i\varphi_L}$ and $\Delta e^{i\varphi_R}$, respectively. The normal region (N) is located at $0<x<L$. Both the longitudinal Josephson current $j_x$ and the transverse charge current $j_y$ can be generated. }
\label{fig:junction}
\end{figure}

In this Letter, we report a study of the transverse transport in the Josephson junctions based on the tilted Dirac materials with valley-dependent gaps. The transverse current is obtained in terms of the Andreev reflection coefficients like the Furusaki-Tsukada formula \cite{FURUSAKI1991299,PhysRevB.73.214511,PhysRevLett.133.226002}, which is analytically derived from the McMillan's formulation \cite{PhysRev.175.559,PhysRevB.56.892}. We find that the finite transverse current appears when the system possesses TRS-broken valley-contrasting gaps. Similar to the conventional longitudinal Josephson current, this transverse current is driven by the superconducting phase difference across the junction and exhibits a nonsinusoidal current-phase relation, leading to a nonreciprocal transverse transport in the Josephson junction, which allows for a larger critical supercurrent in one transverse direction than in the opposite one. This transverse nonreciprocal transport can be termed as the transverse Josephson diode effect (TJDE). The nonreciprocity is present only in the transverse direction and absent in the longitudinal one with a high TJDE efficiency even reaching up to $100\%$, which completely decouples the path of the input signal from that of the output one and suggests potential applications for nonreciprocal superconducting devices. We further analyse the discrete Andreev levels confined in the Josephson junction, which can aid in better understanding the TJDE. It is shown that, as a consequence of the tilted Dirac cones, a pair of asymmetric transverse subgap propagating modes appear for each valley under the broken TRS, which is responsible for the transverse current.

\textit{Model and formalism.}---The Josephson junction under consideration is shown in Fig.\ \ref{fig:junction}, where the longitudinal direction of the junction is along the $x$ axis and the superconducting regions are located at $x<0$ and $x>L$ with $L$ being the length of the junction. The charge carriers in the vicinity of the Dirac points of the anisotropic tilted Dirac systems are described by the Hamiltonian \cite{PhysRevB.103.165415,PhysRevB.102.045417,PhysRevB.105.L201408,PhysRevB.106.165404}
\begin{gather}
    \mathcal{H}_\ell=-i\ell \hbar v_x \partial_x\sigma_x-i\hbar v_y \partial_y\sigma_y-i\ell \hbar v_t\partial_y\sigma_0+\lambda_\ell\sigma_z,
\end{gather}
where $\ell=\pm$ labels $K$ and $K^\prime$ valleys, respectively, $\bm\sigma$ is the Pauli matrix acting on the pseudospin space, $(v_x,v_y)$ and $v_t$ are the direction-dependent velocities and the tilt velocity, respectively. For different valleys, the Dirac cones are tilted in opposite $y$ directions in the momentum space with the same tilt parameter $v_t$. The valley-dependent mass $\lambda_\ell$ is also considered in the normal region of the Josephson junction. Broken TRS leads to the valley-contrasting mass $\lambda_+\neq\lambda_-$, which is responsible for many peculiar phenomena, such as the intriguing optical properties of tilted Dirac systems \cite{PhysRevB.103.165415} and the valley Seebeck effect \cite{PhysRevB.102.045417}. In the superconducting regions, the BCS superconductivity can be induced by the conventional superconductor via the proximity effect \cite{ben2016quantum,PhysRevB.77.205425,PhysRevLett.108.097003}, where the paired electrons are from two different valleys. The Bogoliubov-de Gennes (BdG) Hamiltonian describing the quasiparticles in the Josephson junction can be decoupled into two valley sectors \cite{PhysRevLett.97.067007,RevModPhys.80.1337,de2018superconductivity}. For the $\ell$ valley sector, the BdG Hamiltonian reads
\begin{align}
\mathcal{H}_{BdG}^\ell=\begin{pmatrix}
\mathcal{H}_{\ell}-\mu(x)&\Delta(x)\\
\Delta^\dagger(x)&\mu(x)-\mathcal{H}_{\bar\ell}^*
\end{pmatrix},\label{eq:bdg2}
\end{align}
where the Nambu basis is $(\psi_\ell,\psi^\dagger_{\bar\ell})^T$ with $\psi_\ell$ being the two-component spinors in the sublattice space and $\bar\ell=-\ell$. The chemical potential $\mu(x)$ is taken as $\mu$ in the normal region and $\mu_s$ in the superconducting regions. The pairing potential $\Delta(x)$ is $\Delta e^{i\varphi_L}$ and $\Delta e^{i\varphi_R}$ for left and right superconducting regions, respectively, with $\Delta$ being the amplitude of the pairing potential, $\varphi_L$ and $\varphi_R$ being the corresponding superconducting phases.

In order to obtained the transverse current, we first need the expression for the current operator, which can be derived from the continuity equation. The charge density operator $\hat{\rho}=-e\sum_\ell \psi_\ell^\dagger\psi_\ell$ obeys the Heisenberg equation of motion $\partial_t \hat{\rho}=[\hat{\rho},\mathcal{H}_{BdG}]/i\hbar=-\nabla\cdot\hat{\mathbf{j}}_e-\nabla\cdot\hat{\mathbf{j}}_s$, where $\mathcal{H}_{BdG}$ is the whole BdG Hamiltonian containing both the $K$ and $K'$ valley sectors in Eq.\ (\ref{eq:bdg2}), $\hat{\mathbf{j}}_e$ and $\hat{\mathbf{j}}_s$ are the electronic and source current density operators, respectively. The transverse current operator is obtained after completing the calculation of the commutator, which reads $\hat{j}_y=-e\sum_\ell (v_y\psi_\ell^\dagger\sigma_y\psi_\ell+\ell v_t\psi_\ell^\dagger\psi_\ell)$. The total current density is given by $\mathbf{j}=\langle \hat{\mathbf{j}}_e\rangle+\langle\hat{\mathbf{j}}_s\rangle$, whereas the transverse current is only contributed by the electronic term \cite{Brydon_2013} and can be written as $j_y=\sum_{\ell} j_y^\ell$. Here $j_y^\ell$ is the transverse current density for $\ell$ valley and can be expressed in terms of the Matsubara Green's function:
\begin{align}
    j_y^\ell=-\frac{e}{2\beta}\sum_{k_y,\omega_n}\Big(v_y\mathrm{tr}[\nu_0\sigma_yG_\ell]+\ell v_t\mathrm{tr}[\nu_0\sigma_0G_\ell]\Big),\label{eq:jyy}
\end{align}
where $\beta$ is the inverse temperature, $G_\ell=(i\omega_n-H_{BdG}^\ell)^{-1}$ is the Matsubara Green's function for the $\ell$ sector of the BdG Hamiltonian in Eq.\ (\ref{eq:bdg2}) with $\omega_n=(2n+1)\pi/\beta$, ($n=0,\pm1,\pm2\dotsc$) and can be obtained by the analytical continuation from the retarded Green's function $G_\ell^R$. 

\begin{figure}[tp]
\begin{center}
\includegraphics[clip = true, width =\columnwidth]{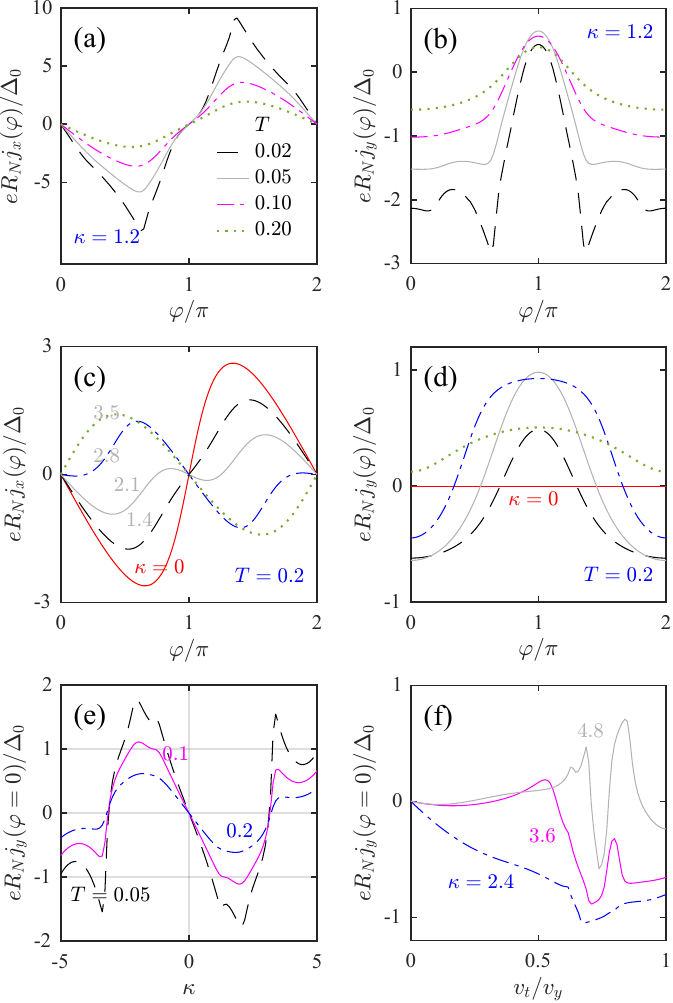}
\end{center}
\caption{[(a), (b)] Current-phase relation for $j_x(\varphi)$ and $j_y(\varphi)$ at $\kappa=1.2$. [(c), (d)] Current-phase relation for $j_x(\varphi)$ and $j_y(\varphi)$ at $T=0.2$ (in units of $T_c$). (e) Transverse current at zero phase difference [$j_y(\varphi=0)$] as a function of $\kappa$. (f) $j_y(\varphi=0)$ as a function of the tilt parameter $v_t$. The other parameters are $\mu_s=200\Delta_0$ and $\mu L/\hbar v_F=4.2$. We normalize the current to $eR_Nj_{x,y}/\Delta_0$ with $R_N$ being the resistivity of the junction in normal state.}
\label{fig:cpr}
\end{figure}

We follow the McMillan's approach \cite{PhysRev.175.559,PhysRevB.56.892,FURUSAKI1991299} for constructing the retarded Green's function $G_\ell^R$; see Supplemental Material. Because of the translational invariance along the $y$ direction, the retarded Green's function can be written as $G_\ell^R(\mathbf{r},\mathbf{r}^\prime)=G_{\ell}^{R}(x,x')e^{ik_y(y-y')}$ with $k_y$ being the conserved transverse wave number and  
\begin{align}
    G_{\ell}^{R}(x,x')=\varrho\times\left\{
        \begin{array}{ll}
         \Xi_1^\ell/\cos\theta_1-\Xi_2^\ell/\cos\theta_2, & x>x', \\
          \tilde\Xi_1^\ell/\cos\theta_1-\tilde\Xi_2^\ell/\cos\theta_2, & x<x',
        \end{array}
    \right.\label{eq:gf0}
\end{align}
where $\varrho=-iv_x^{-1}\Omega/(E+\Omega)$, $\Omega=\sqrt{E^2-\Delta^2}$ for $E>\Delta$ and $\Omega=i\sqrt{\Delta^2-E^2}$ for $E<\Delta$ with $E$ being the energy measured from the Fermi level. $\Xi_{1,2}^\ell$ and $\tilde\Xi_{1,2}^\ell$ are matrices constructed by the scattering basis spinors of $H_{BdG}^\ell$ with $\theta_{1,2}$ being the corresponding transmission angles, which are provided in Supplemental Material. The Matsubara Green's function can be obtained by the analytic continuation $E+i 0^+\rightarrow i\omega_n$ from the retarded Green's function in Eq.\ (\ref{eq:gf0}). We consider the short junction $L\ll\xi_0$, where $\xi_0=\hbar v_F/\Delta$ is the superconducting coherence length. Without loss of generality, we assume $\varphi_L=0$ and $\varphi_R=\varphi$ in our model and set $\hbar=1$ throughout our calculation. The mean-field treatment of superconductivity requires the heavy-doping limit in the superconducting regions, \textit{i.e.}, $|\mu_s|\gg\{|\mu|,\Delta\}$ \cite{PhysRevLett.97.067007,RevModPhys.80.1337}. In this regime, the transverse current density is expressed by the Andreev reflection coefficients
\begin{align}
  j_{y}^\ell=&ek_BT\sum_{k_y,\omega_n}\frac{\Delta}{\sqrt{\omega_n^2+\Delta^2}}\times\frac{v_y\tan\theta+\ell v_t\sec\theta}{v_x}\nonumber\\&\times\mathfrak{Re}\Big(a^\ell_1(\varphi,i\omega_n)+a^\ell_2(\varphi,i\omega_n)\Big)e^{2\Omega_nx/v_x},\label{eq:jjy}
\end{align}
where $k_B$ is the Boltzmann constant, $T$ is the temperature, $\theta=\arctan(v_yk_y/\mu_s)$, $\Omega_n=\sqrt{\omega_n^2+\Delta^2}$, $a^\ell_1(\varphi,i\omega_n)$ and $a^\ell_2(\varphi,i\omega_n)$ are the Andreev reflection amplitudes for the electron-like quasiparticles and hole-like quasiparticles, respectively. The longitudinal current (Josephson current) is given by $j_x=\sum_\ell j_x^\ell$, where $j_x^\ell$ is obtained from the Furusaki-Tsukada formula \cite{FURUSAKI1991299,PhysRevB.73.214511,PhysRevLett.133.226002}
\begin{align}
  j_{x}^\ell=ek_BT\sum_{k_y,\omega_n}\frac{\Delta}{\sqrt{\omega_n^2+\Delta^2}}\Big(a^\ell_1(\varphi,i\omega_n)-a^\ell_2(\varphi,i\omega_n)\Big).\label{eq:jjx}
\end{align}

\textit{Transverse Josephson diode effect.}---We consider the parameters $v_x=0.86v_F$, $v_y=0.69v_F$, and $v_t=0.32v_F$ corresponding to the two-dimensional material $8$-\textit{Pmmn} borophene \cite{PhysRevB.93.241405,PhysRevB.99.155418,PhysRevB.109.205413,PhysRevB.99.035415}, where $v_F=\SI[parse-numbers=false]{10^6}{\meter/\second}$. The temperature dependence of the pairing potential is assumed to be $\Delta(T)=\Delta_0\mathrm{tanh}(1.74\sqrt{T_c/T-1})$ with $\Delta_0=1.76k_BT_c$, where $T_c$ is the critical temperature. We define a dimensionless parameter, $\kappa=(\lambda_+-\lambda_-)/\Delta_0$, to represent the discrepancy between the two valley-dependent gaps.

The current-phase relation (CPR) of the longitudinal Josephson current density $j_x(\varphi)$ is shown in Fig.\ \ref{fig:cpr}(a) at $\kappa=1.2$. $j_x(\varphi)$ exhibits a sinusoidal $\pi$-phase CPR. The critical current of $j_x(\varphi)$ decreases with the increasing the temperature $T$, but the CPR remains a $\pi$ phase. $j_x(\varphi)$ at $T=0.2$ (in units of $T_c$) for different $\kappa$ is shown in Fig.\ \ref{fig:cpr}(c). The Josephson current changes its sign and the $0$-$\pi$ phase transition occurs with the increasing of $\kappa$. Similar to the $0$-$\pi$ transition in SFS Josephson junctions \cite{PhysRevB.83.144515,PhysRevB.76.134502} (F represents the ferromagnetic metal), the supercurrent reversal is attributed to the valley polarization induced by the broken TRS, where the valley-singlet Cooper pairs acquire a nonzero momentum \cite{PhysRevB.89.064501}. The Josephson current holds the relation $j_x(\varphi)=-j_x(-\varphi)$ for both $0$- and $\pi$-phase CPRs. Consequently, the anomalous supercurrent flowing at zero phase difference is absent. The forward and backward critical currents are symmetric, \textit{i.e.}, $j_x^+=j_x^-$, which indicates the absence of the diode effect in the longitudinal direction. 

The transverse current has its maximum value at the NS boundary and decays exponentially into the S region, we present the CPRs of the transverse current at the NS interface ($x=0$) in Figs.\ \ref{fig:cpr}(b) and \ref{fig:cpr}(d). The critical current of $j_y(\varphi)$ decreases with the increasing of $T$; see Fig.\ \ref{fig:cpr}(b). Distinct from the longitudinal Josephson current, $j_y(\varphi)$ disappears at $\kappa=0$ where the TRS is preserved, as shown in Fig.\ \ref{fig:cpr}(d). However, a finite transverse current flows even at zero phase difference. This anomalous transverse current $j_y(\varphi=0)$ exhibits a symmetric relation of $\kappa$ [$j_y(\varphi=0,\kappa)=-j_y(\varphi=0,-\kappa)$] and can be further enhanced by either decreasing the temperature or tilting the Dirac cones; see Figs.\ \ref{fig:cpr}(e) and \ref{fig:cpr}(f). 

Analogous to the Hall angle of the conventional tunneling Hall effect in tunnel junctions \cite{PhysRevLett.131.246301,PhysRevB.110.024511,PhysRevLett.115.056602,PhysRevLett.117.166806}, the efficiency of the conversion of the longitudinal and transverse current in the Josephson junction is in proportion to $j_y/j_x$, reaching its maximum value at $\varphi=n\pi$, ($n\in\mathbb{Z}$), where a pure transverse current is present without the longitudinal Josephson current. There is a subtle difference between the predicted Josephson Hall effect in our model and the previously proposed tunneling Hall effect in the tunnel junctions based on the tilted Dirac materials. The latter effect is associated with the momentum filtering due to the mismatch of the tilted Fermi surfaces in different regions of the heterojunctions \cite{PhysRevLett.131.246301,PhysRevB.110.024511}, where the electron tunneling is skewed. However, we will point out that this skew-tunneling mechanism plays a minor role in the Josephson Hall effect. The transverse current in Eq.\ (\ref{eq:jyy}) can be decomposed into two parts: $j_y^\ell=j_{y,1}^{\ell}+j_{y,2}^{\ell}$ with $j_{y,1}^{\ell}\sim (v_y/v_x)\sum_{j=1,2}\int_{-\pi/2}^{\pi/2}d\theta a_{j}^\ell(\theta)\tan\theta$ and $j_{y,2}^{\ell}\sim (v_t/v_x)\sum_{j=1,2}\int_{-\pi/2}^{\pi/2}d\theta \ell a_{j}^\ell(\theta)/\cos\theta$, where we replace the summation over $k_y$ by the integration over $\theta$. $j_{y,1}^\ell$ is related to the skew-tunneling and is absent when the scattering is symmetric, \textit{i.e.}, $a_{j}^\ell(\theta)=a_{j}^\ell(-\theta)$, whereas $j_{y,2}^\ell$ is attributed to the tilt-induced nonzero transverse momentum acquired by the quasiparticles. In the heavy-doping regime, $j_{y,1}^\ell$ is much smaller than $j_{y,2}^\ell$ with the scale $j_{y,1}^\ell/j_{y,2}^\ell\sim|\mu/\mu_s|$. As a result, $j_{y,2}^\ell$ is dominant in the proposed Josephson Hall effect, where the transverse current is driven by the passage of the Cooper pair across the junction carrying a finite tilt-induced transverse momentum.

\begin{figure}[tp]
\begin{center}
\includegraphics[clip = true, width =\columnwidth]{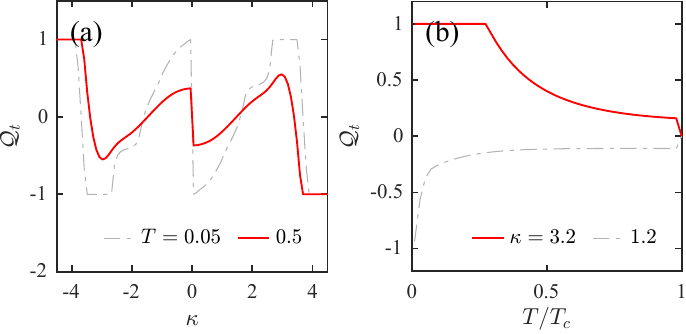}
\end{center}
\caption{(a) Quality factor of the TJDE $\mathcal{Q}_t$ as a function of $\kappa$. (b) Quality factor versus the temperature $T$. }
\label{fig:qt}
\end{figure}

Most notably, $j_y(\varphi)$ is a periodic function of $\varphi$ with a period of $2\pi$ but possesses a nonsinusoidal CPR, which indicates a nonreciprocal transverse charge transport in the Josephson junction. The extent of the nonreciprocal in the transverse direction of the Josephson junction is characterized by the quality factor:
\begin{align}
    \mathcal{Q}_t=\frac{j_{y,c}^+-|j_{y,c}^-|}{j_{y,c}^++|j_{y,c}^-|},
\end{align}
where $j_{y,c}^+$ ($j_{y,c}^-$) is the critical current along the positive (negative) transverse direction of the junction and the sign of $\mathcal{Q}_t$ defines the transverse diode polarity. In Fig.\ \ref{fig:qt}(a), we present the quality factor of the TJDE as a function of $\kappa$. The quality factor exhibits the symmetric relation $\mathcal{Q}_t(\kappa)=-\mathcal{Q}_t(-\kappa)$. This symmetric behavior of $\mathcal{Q}_t$ is similar to that of the longitudinal Josephson diode effect predicted on the surface of topological insulators \cite{PhysRevLett.131.096001}, where the broken TRS is induced by the external magnetic field. Remarkably, the quality factor can even reach $100\%$ by tuning the valley-dependent band gaps. The temperature-dependence of the transverse quality factor is shown in Fig.\ \ref{fig:qt}(b). For both the positive and negative transverse diode polarity [$\mathcal{Q}_t>0$ and $\mathcal{Q}_t<0$, respectively], the TJDE efficiency at low temperature is generally greater than that at high temperature.

Although the transverse current obtained in Eq.\ (\ref{eq:jjy}) includes contributions from both the discrete and continuum spectra, the TJDE can be qualitatively understood in terms of the Andreev bound states (ABSs) confined in the normal region, which arise from the closed-loop motion of the carriers with discrete subgap energies. For $E<\Delta$, there are no propagating modes in the superconducting regions. We can then define the transfer matrix for the Andreev reflection at the NS interfaces, which is given by $\mathcal{S}_{+(-)}=e^{-i\varphi_{R(L)}-(+)i\ell\alpha\sigma_x}$ with $\alpha=\arccos(E/\Delta)$ for the right (left) interface \cite{PhysRevB.74.041401,PhysRevB.105.094510}. The transfer matrix for the carriers in the normal region can be obtained by the direct integration of the BdG equation \cite{PhysRevB.77.085423,PhysRevB.79.205428}, which yields $\psi(x)=e^{\int_{x_0}^x\ i\ell v_x^{-1}\nu_z\sigma_x\mathcal{O}dx'}\psi(x_0)$ with $\mathcal{O}=E\times\mathds{1}-\mathcal{H}_{BdG}^\ell|_{k_x=0}$. The electron and hole states are decoupled in the normal region due to the absence of the superconductivity. As a result, $\mathcal{O}$ is a block diagonal  matrix. Evaluating the integral across the normal region gives rise to
\begin{align}
  e^{i\ell L\nu_z\sigma_x\mathcal{O}/v_x}=\begin{pmatrix}
    \mathcal{M}&\slashzero\\
    \slashzero&\mathcal{M}^\prime
  \end{pmatrix},
\end{align}
where $\mathcal{M}$ and $\mathcal{M}^\prime$ are the transfer matrices for the electron and hole states, respectively. For the existence of ABSs in the Josephson junction, the Andreev process between two NS interfaces forms a closed loop and the transfer matrix $\mathcal{M}^{-1}\mathcal{S}_{+}^{-1}\mathcal{M}^{\prime}\mathcal{S}_{-}$ is unimodular, which in turn gives the discrete Andreev levels $E_{ABS}$.

\begin{figure}[tp]
\begin{center}
\includegraphics[clip = true, width =\columnwidth]{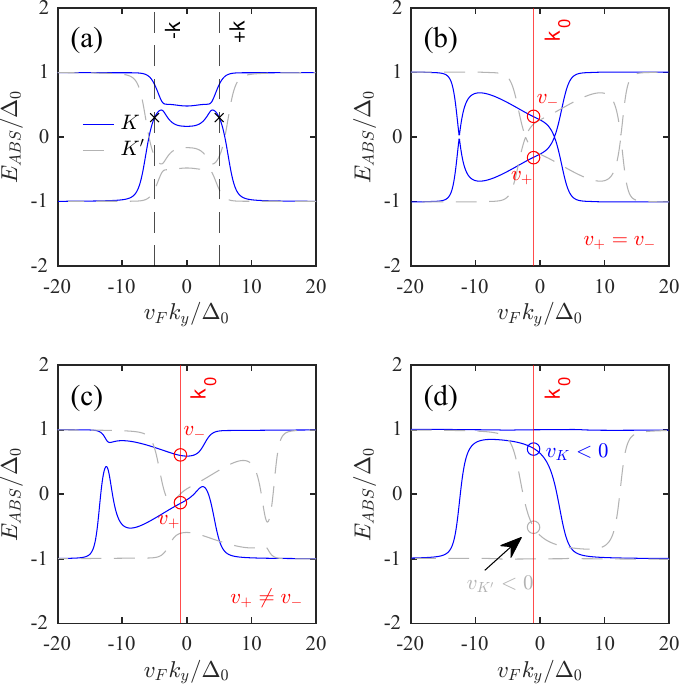}
\end{center}
\caption{Andreev level $E_{ABS}$ versus the transverse wave number $k_y$ at $T=\qty{0}{\kelvin}$. The blue solid lines and the gray dashed lines denote the $E_{ABS}$ for $K$ and $K'$ valleys, respectively. (a) $v_t=0$ and $\kappa=1.2$. (b) $v_t=0.32v_F$ and $\kappa=0$. [(c), (d)] $v_t=0.32v_F$, $\kappa=1.2$ and $1.6$ for (c) and (d), respectively. The other parameters are $\varphi_L=0$, $\varphi_R=\pi$, and $\mu_s=200\Delta_0$.}
\label{fig:abs}
\end{figure}

The numerical results of the transverse-momentum-dependence of $E_{ABS}$ are presented in Fig.\ \ref{fig:abs}. When the tilt is absent ($v_t=0$), the ABS energy levels are symmetric with respect to the transverse momentum, \textit{i.e.}, $E_{ABS}(k_y,\varphi)=E_{ABS}(-k_y,\varphi)$; see Fig.\ \ref{fig:abs}(a). For each transverse propagating mode at $|+k_y\rangle$ with the group velocity $v(k_y)=\partial E_{ABS}(k_y,\varphi)/\partial k_y$, there always exists an oppositely propagating mode at $|-k_y\rangle$ with its group velocity being $-v(k_y)$. Consequently, the transverse current is always absent.

For the TRS-preserved tilted Dirac systems ($\kappa=0$ and $v_t\neq0$), the ABS spectra are skewed to opposite $k_y$ directions for the opposite valleys due to the tilt of the Dirac cones; see Fig.\ \ref{fig:abs}(b). However, $E_{ABS}$ is symmetric with respect to $E=0$ due to TRS, which holds the relation $ E_{ABS}(k_y,\varphi)=-E_{ABS}(k_y,\varphi)$. For each transverse propagating mode with group velocity $v_+$, there always exists an oppositely propagating mode at the same transverse momentum with group velocity $v_-$, where $v_+=v_-$. Thus, the transverse current is also absent.

The ABS energy levels of the broken-TRS tilted Dirac systems are shown in Figs.\ \ref{fig:abs}(c) and \ref{fig:abs}(d). We focus on the small $k_y$ regime, which makes a dominant contribution in Josephson effects. For a given valley, there are two subgap transverse modes propagating in the opposite $y$ directions at $\kappa=1.2$, with their group velocities being $v_+$ and $v_-$, respectively; see Fig.\ \ref{fig:abs}(c). The transverse nonreciprocal transport is generated as a consequence of $v_+\neq v_-$. For $\kappa=1.6$, only one propagating subgap mode appears for each valley; see Fig.\ \ref{fig:abs}(d). In the small $k_y$ regime, both the subgap modes from $K$ and $K'$ valley are propagating along the negative transverse direction with their group velocities being $v_{K,K'}<0$. The positive propagating modes are absent. Distinct from the conventional Josephson diode effect along the longitudinal direction, the predicted TJDE is attributed to the tilt-induced asymmetric transverse propagating modes, rather than the Doppler shift of the quasiparticle energies \cite{dsp1,PhysRevLett.128.037001,PhysRevApplied.21.054057}.

\textit{Conclusions.}---To conclude, we study the transverse charge transport in the Josephson junctions based on the tilted Dirac materials with valley-dependent gaps. The transverse charge current is expressed by Andreev reflection coefficients like the Furusaki-Tsukada formula, which is theoretically obtained by using the McMillan's Green's function. It is shown that the tilt-assisted transverse current can be generated under broken TRS and is driven by the superconducting phase difference across the junction. The transverse critical current is direction-selective, resulting in a nonreciprocal transverse transport, namely, the transverse Josephson diode effect. The nonreciprocity is present only in the transverse direction and absent in the longitudinal one with a high TJDE efficiency even reaching up to $100\%$, which completely decouples the path of the input signal from that of the output one and suggests potential applications for nonreciprocal superconducting devices.

\textit{Acknowledgements.}---I wish gratefully to acknowledge discussions with Professor Bo Lu (Tianjin University) and Professor Wei Chen (Nanjing University).


\begin{thebibliography}{60}%
\makeatletter
\providecommand \@ifxundefined [1]{%
 \@ifx{#1\undefined}
}%
\providecommand \@ifnum [1]{%
 \ifnum #1\expandafter \@firstoftwo
 \else \expandafter \@secondoftwo
 \fi
}%
\providecommand \@ifx [1]{%
 \ifx #1\expandafter \@firstoftwo
 \else \expandafter \@secondoftwo
 \fi
}%
\providecommand \natexlab [1]{#1}%
\providecommand \enquote  [1]{``#1''}%
\providecommand \bibnamefont  [1]{#1}%
\providecommand \bibfnamefont [1]{#1}%
\providecommand \citenamefont [1]{#1}%
\providecommand \href@noop [0]{\@secondoftwo}%
\providecommand \href [0]{\begingroup \@sanitize@url \@href}%
\providecommand \@href[1]{\@@startlink{#1}\@@href}%
\providecommand \@@href[1]{\endgroup#1\@@endlink}%
\providecommand \@sanitize@url [0]{\catcode `\\12\catcode `\$12\catcode
  `\&12\catcode `\#12\catcode `\^12\catcode `\_12\catcode `\%12\relax}%
\providecommand \@@startlink[1]{}%
\providecommand \@@endlink[0]{}%
\providecommand \url  [0]{\begingroup\@sanitize@url \@url }%
\providecommand \@url [1]{\endgroup\@href {#1}{\urlprefix }}%
\providecommand \urlprefix  [0]{URL }%
\providecommand \Eprint [0]{\href }%
\providecommand \doibase [0]{https://doi.org/}%
\providecommand \selectlanguage [0]{\@gobble}%
\providecommand \bibinfo  [0]{\@secondoftwo}%
\providecommand \bibfield  [0]{\@secondoftwo}%
\providecommand \translation [1]{[#1]}%
\providecommand \BibitemOpen [0]{}%
\providecommand \bibitemStop [0]{}%
\providecommand \bibitemNoStop [0]{.\EOS\space}%
\providecommand \EOS [0]{\spacefactor3000\relax}%
\providecommand \BibitemShut  [1]{\csname bibitem#1\endcsname}%
\let\auto@bib@innerbib\@empty
%</preamble>
\bibitem [{\citenamefont {Zhang}\ \emph {et~al.}(2022)\citenamefont {Zhang},
  \citenamefont {Gu}, \citenamefont {Li}, \citenamefont {Hu},\ and\
  \citenamefont {Jiang}}]{PhysRevX.12.041013}%
  \BibitemOpen
  \bibfield  {author} {\bibinfo {author} {\bibfnamefont {Y.}~\bibnamefont
  {Zhang}}, \bibinfo {author} {\bibfnamefont {Y.}~\bibnamefont {Gu}}, \bibinfo
  {author} {\bibfnamefont {P.}~\bibnamefont {Li}}, \bibinfo {author}
  {\bibfnamefont {J.}~\bibnamefont {Hu}},\ and\ \bibinfo {author}
  {\bibfnamefont {K.}~\bibnamefont {Jiang}},\ }\bibfield  {title} {\bibinfo
  {title} {General theory of {Josephson} diodes},\ }\href
  {https://doi.org/10.1103/PhysRevX.12.041013} {\bibfield  {journal} {\bibinfo
  {journal} {Phys. Rev. X}\ }\textbf {\bibinfo {volume} {12}},\ \bibinfo
  {pages} {041013} (\bibinfo {year} {2022})}\BibitemShut {NoStop}%
\bibitem [{\citenamefont {Hu}\ \emph {et~al.}(2007)\citenamefont {Hu},
  \citenamefont {Wu},\ and\ \citenamefont {Dai}}]{PhysRevLett.99.067004}%
  \BibitemOpen
  \bibfield  {author} {\bibinfo {author} {\bibfnamefont {J.}~\bibnamefont
  {Hu}}, \bibinfo {author} {\bibfnamefont {C.}~\bibnamefont {Wu}},\ and\
  \bibinfo {author} {\bibfnamefont {X.}~\bibnamefont {Dai}},\ }\bibfield
  {title} {\bibinfo {title} {Proposed design of a {Josephson} diode},\ }\href
  {https://doi.org/10.1103/PhysRevLett.99.067004} {\bibfield  {journal}
  {\bibinfo  {journal} {Phys. Rev. Lett.}\ }\textbf {\bibinfo {volume} {99}},\
  \bibinfo {pages} {067004} (\bibinfo {year} {2007})}\BibitemShut {NoStop}%
\bibitem [{\citenamefont {Liu}\ \emph {et~al.}(2024)\citenamefont {Liu},
  \citenamefont {Huang},\ and\ \citenamefont {Wang}}]{PhysRevB.110.014519}%
  \BibitemOpen
  \bibfield  {author} {\bibinfo {author} {\bibfnamefont {Z.}~\bibnamefont
  {Liu}}, \bibinfo {author} {\bibfnamefont {L.}~\bibnamefont {Huang}},\ and\
  \bibinfo {author} {\bibfnamefont {J.}~\bibnamefont {Wang}},\ }\bibfield
  {title} {\bibinfo {title} {Josephson diode effect in topological
  superconductors},\ }\href {https://doi.org/10.1103/PhysRevB.110.014519}
  {\bibfield  {journal} {\bibinfo  {journal} {Phys. Rev. B}\ }\textbf {\bibinfo
  {volume} {110}},\ \bibinfo {pages} {014519} (\bibinfo {year}
  {2024})}\BibitemShut {NoStop}%
\bibitem [{\citenamefont {Hu}\ \emph {et~al.}(2023)\citenamefont {Hu},
  \citenamefont {Sun}, \citenamefont {Xie},\ and\ \citenamefont
  {Law}}]{PhysRevLett.130.266003}%
  \BibitemOpen
  \bibfield  {author} {\bibinfo {author} {\bibfnamefont {J.-X.}\ \bibnamefont
  {Hu}}, \bibinfo {author} {\bibfnamefont {Z.-T.}\ \bibnamefont {Sun}},
  \bibinfo {author} {\bibfnamefont {Y.-M.}\ \bibnamefont {Xie}},\ and\ \bibinfo
  {author} {\bibfnamefont {K.~T.}\ \bibnamefont {Law}},\ }\bibfield  {title}
  {\bibinfo {title} {Josephson diode effect induced by valley polarization in
  twisted bilayer graphene},\ }\href
  {https://doi.org/10.1103/PhysRevLett.130.266003} {\bibfield  {journal}
  {\bibinfo  {journal} {Phys. Rev. Lett.}\ }\textbf {\bibinfo {volume} {130}},\
  \bibinfo {pages} {266003} (\bibinfo {year} {2023})}\BibitemShut {NoStop}%
\bibitem [{\citenamefont {Souto}\ \emph {et~al.}(2022)\citenamefont {Souto},
  \citenamefont {Leijnse},\ and\ \citenamefont
  {Schrade}}]{PhysRevLett.129.267702}%
  \BibitemOpen
  \bibfield  {author} {\bibinfo {author} {\bibfnamefont {R.~S.}\ \bibnamefont
  {Souto}}, \bibinfo {author} {\bibfnamefont {M.}~\bibnamefont {Leijnse}},\
  and\ \bibinfo {author} {\bibfnamefont {C.}~\bibnamefont {Schrade}},\
  }\bibfield  {title} {\bibinfo {title} {Josephson diode effect in supercurrent
  interferometers},\ }\href {https://doi.org/10.1103/PhysRevLett.129.267702}
  {\bibfield  {journal} {\bibinfo  {journal} {Phys. Rev. Lett.}\ }\textbf
  {\bibinfo {volume} {129}},\ \bibinfo {pages} {267702} (\bibinfo {year}
  {2022})}\BibitemShut {NoStop}%
\bibitem [{\citenamefont {Lu}\ \emph {et~al.}(2023)\citenamefont {Lu},
  \citenamefont {Ikegaya}, \citenamefont {Burset}, \citenamefont {Tanaka},\
  and\ \citenamefont {Nagaosa}}]{PhysRevLett.131.096001}%
  \BibitemOpen
  \bibfield  {author} {\bibinfo {author} {\bibfnamefont {B.}~\bibnamefont
  {Lu}}, \bibinfo {author} {\bibfnamefont {S.}~\bibnamefont {Ikegaya}},
  \bibinfo {author} {\bibfnamefont {P.}~\bibnamefont {Burset}}, \bibinfo
  {author} {\bibfnamefont {Y.}~\bibnamefont {Tanaka}},\ and\ \bibinfo {author}
  {\bibfnamefont {N.}~\bibnamefont {Nagaosa}},\ }\bibfield  {title} {\bibinfo
  {title} {Tunable {Josephson} diode effect on the surface of topological
  insulators},\ }\href {https://doi.org/10.1103/PhysRevLett.131.096001}
  {\bibfield  {journal} {\bibinfo  {journal} {Phys. Rev. Lett.}\ }\textbf
  {\bibinfo {volume} {131}},\ \bibinfo {pages} {096001} (\bibinfo {year}
  {2023})}\BibitemShut {NoStop}%
\bibitem [{\citenamefont {Wang}\ \emph {et~al.}(2024)\citenamefont {Wang},
  \citenamefont {Jiang}, \citenamefont {Wang},\ and\ \citenamefont
  {Liu}}]{PhysRevB.109.075412}%
  \BibitemOpen
  \bibfield  {author} {\bibinfo {author} {\bibfnamefont {J.}~\bibnamefont
  {Wang}}, \bibinfo {author} {\bibfnamefont {Y.}~\bibnamefont {Jiang}},
  \bibinfo {author} {\bibfnamefont {J.~J.}\ \bibnamefont {Wang}},\ and\
  \bibinfo {author} {\bibfnamefont {J.-F.}\ \bibnamefont {Liu}},\ }\bibfield
  {title} {\bibinfo {title} {Efficient {Josephson} diode effect on a
  two-dimensional topological insulator with asymmetric magnetization},\ }\href
  {https://doi.org/10.1103/PhysRevB.109.075412} {\bibfield  {journal} {\bibinfo
   {journal} {Phys. Rev. B}\ }\textbf {\bibinfo {volume} {109}},\ \bibinfo
  {pages} {075412} (\bibinfo {year} {2024})}\BibitemShut {NoStop}%
\bibitem [{\citenamefont {Costa}\ \emph {et~al.}(2023)\citenamefont {Costa},
  \citenamefont {Fabian},\ and\ \citenamefont {Kochan}}]{PhysRevB.108.054522}%
  \BibitemOpen
  \bibfield  {author} {\bibinfo {author} {\bibfnamefont {A.}~\bibnamefont
  {Costa}}, \bibinfo {author} {\bibfnamefont {J.}~\bibnamefont {Fabian}},\ and\
  \bibinfo {author} {\bibfnamefont {D.}~\bibnamefont {Kochan}},\ }\bibfield
  {title} {\bibinfo {title} {Microscopic study of the {Josephson} supercurrent
  diode effect in {Josephson} junctions based on two-dimensional electron
  gas},\ }\href {https://doi.org/10.1103/PhysRevB.108.054522} {\bibfield
  {journal} {\bibinfo  {journal} {Phys. Rev. B}\ }\textbf {\bibinfo {volume}
  {108}},\ \bibinfo {pages} {054522} (\bibinfo {year} {2023})}\BibitemShut
  {NoStop}%
\bibitem [{\citenamefont {Cayao}\ \emph {et~al.}(2024)\citenamefont {Cayao},
  \citenamefont {Nagaosa},\ and\ \citenamefont
  {Tanaka}}]{PhysRevB.109.L081405}%
  \BibitemOpen
  \bibfield  {author} {\bibinfo {author} {\bibfnamefont {J.}~\bibnamefont
  {Cayao}}, \bibinfo {author} {\bibfnamefont {N.}~\bibnamefont {Nagaosa}},\
  and\ \bibinfo {author} {\bibfnamefont {Y.}~\bibnamefont {Tanaka}},\
  }\bibfield  {title} {\bibinfo {title} {Enhancing the {Josephson} diode effect
  with {Majorana} bound states},\ }\href
  {https://doi.org/10.1103/PhysRevB.109.L081405} {\bibfield  {journal}
  {\bibinfo  {journal} {Phys. Rev. B}\ }\textbf {\bibinfo {volume} {109}},\
  \bibinfo {pages} {L081405} (\bibinfo {year} {2024})}\BibitemShut {NoStop}%
\bibitem [{\citenamefont {Volkov}\ \emph {et~al.}(2024)\citenamefont {Volkov},
  \citenamefont {Lantagne-Hurtubise}, \citenamefont {Tummuru}, \citenamefont
  {Plugge}, \citenamefont {Pixley},\ and\ \citenamefont
  {Franz}}]{PhysRevB.109.094518}%
  \BibitemOpen
  \bibfield  {author} {\bibinfo {author} {\bibfnamefont {P.~A.}\ \bibnamefont
  {Volkov}}, \bibinfo {author} {\bibfnamefont {E.}~\bibnamefont
  {Lantagne-Hurtubise}}, \bibinfo {author} {\bibfnamefont {T.}~\bibnamefont
  {Tummuru}}, \bibinfo {author} {\bibfnamefont {S.}~\bibnamefont {Plugge}},
  \bibinfo {author} {\bibfnamefont {J.~H.}\ \bibnamefont {Pixley}},\ and\
  \bibinfo {author} {\bibfnamefont {M.}~\bibnamefont {Franz}},\ }\bibfield
  {title} {\bibinfo {title} {Josephson diode effects in twisted nodal
  superconductors},\ }\href {https://doi.org/10.1103/PhysRevB.109.094518}
  {\bibfield  {journal} {\bibinfo  {journal} {Phys. Rev. B}\ }\textbf {\bibinfo
  {volume} {109}},\ \bibinfo {pages} {094518} (\bibinfo {year}
  {2024})}\BibitemShut {NoStop}%
\bibitem [{\citenamefont {Scharf}\ \emph {et~al.}(2024)\citenamefont {Scharf},
  \citenamefont {Kochan},\ and\ \citenamefont
  {Matos-Abiague}}]{PhysRevB.110.134511}%
  \BibitemOpen
  \bibfield  {author} {\bibinfo {author} {\bibfnamefont {B.}~\bibnamefont
  {Scharf}}, \bibinfo {author} {\bibfnamefont {D.}~\bibnamefont {Kochan}},\
  and\ \bibinfo {author} {\bibfnamefont {A.}~\bibnamefont {Matos-Abiague}},\
  }\bibfield  {title} {\bibinfo {title} {Superconducting diode effect in
  quantum spin {Hall} insulator based {Josephson} junctions},\ }\href
  {https://doi.org/10.1103/PhysRevB.110.134511} {\bibfield  {journal} {\bibinfo
   {journal} {Phys. Rev. B}\ }\textbf {\bibinfo {volume} {110}},\ \bibinfo
  {pages} {134511} (\bibinfo {year} {2024})}\BibitemShut {NoStop}%
\bibitem [{\citenamefont {Tanaka}\ \emph {et~al.}(2022)\citenamefont {Tanaka},
  \citenamefont {Lu},\ and\ \citenamefont {Nagaosa}}]{PhysRevB.106.214524}%
  \BibitemOpen
  \bibfield  {author} {\bibinfo {author} {\bibfnamefont {Y.}~\bibnamefont
  {Tanaka}}, \bibinfo {author} {\bibfnamefont {B.}~\bibnamefont {Lu}},\ and\
  \bibinfo {author} {\bibfnamefont {N.}~\bibnamefont {Nagaosa}},\ }\bibfield
  {title} {\bibinfo {title} {Theory of giant diode effect in $d$-wave
  superconductor junctions on the surface of a topological insulator},\ }\href
  {https://doi.org/10.1103/PhysRevB.106.214524} {\bibfield  {journal} {\bibinfo
   {journal} {Phys. Rev. B}\ }\textbf {\bibinfo {volume} {106}},\ \bibinfo
  {pages} {214524} (\bibinfo {year} {2022})}\BibitemShut {NoStop}%
\bibitem [{\citenamefont {Fu}\ \emph {et~al.}(2024)\citenamefont {Fu},
  \citenamefont {Xu}, \citenamefont {Yang}, \citenamefont {Lee}, \citenamefont
  {Ang},\ and\ \citenamefont {Liu}}]{PhysRevApplied.21.054057}%
  \BibitemOpen
  \bibfield  {author} {\bibinfo {author} {\bibfnamefont {P.-H.}\ \bibnamefont
  {Fu}}, \bibinfo {author} {\bibfnamefont {Y.}~\bibnamefont {Xu}}, \bibinfo
  {author} {\bibfnamefont {S.~A.}\ \bibnamefont {Yang}}, \bibinfo {author}
  {\bibfnamefont {C.~H.}\ \bibnamefont {Lee}}, \bibinfo {author} {\bibfnamefont
  {Y.~S.}\ \bibnamefont {Ang}},\ and\ \bibinfo {author} {\bibfnamefont {J.-F.}\
  \bibnamefont {Liu}},\ }\bibfield  {title} {\bibinfo {title} {Field-effect
  {Josephson} diode via asymmetric spin-momentum locking states},\ }\href
  {https://doi.org/10.1103/PhysRevApplied.21.054057} {\bibfield  {journal}
  {\bibinfo  {journal} {Phys. Rev. Appl.}\ }\textbf {\bibinfo {volume} {21}},\
  \bibinfo {pages} {054057} (\bibinfo {year} {2024})}\BibitemShut {NoStop}%
\bibitem [{\citenamefont {Chiles}\ \emph {et~al.}(2023)\citenamefont {Chiles},
  \citenamefont {Arnault}, \citenamefont {Chen}, \citenamefont {Larson},
  \citenamefont {Zhao}, \citenamefont {Watanabe}, \citenamefont {Taniguchi},
  \citenamefont {Amet},\ and\ \citenamefont
  {Finkelstein}}]{chiles2023nonreciprocal}%
  \BibitemOpen
  \bibfield  {author} {\bibinfo {author} {\bibfnamefont {J.}~\bibnamefont
  {Chiles}}, \bibinfo {author} {\bibfnamefont {E.~G.}\ \bibnamefont {Arnault}},
  \bibinfo {author} {\bibfnamefont {C.-C.}\ \bibnamefont {Chen}}, \bibinfo
  {author} {\bibfnamefont {T.~F.}\ \bibnamefont {Larson}}, \bibinfo {author}
  {\bibfnamefont {L.}~\bibnamefont {Zhao}}, \bibinfo {author} {\bibfnamefont
  {K.}~\bibnamefont {Watanabe}}, \bibinfo {author} {\bibfnamefont
  {T.}~\bibnamefont {Taniguchi}}, \bibinfo {author} {\bibfnamefont
  {F.}~\bibnamefont {Amet}},\ and\ \bibinfo {author} {\bibfnamefont
  {G.}~\bibnamefont {Finkelstein}},\ }\bibfield  {title} {\bibinfo {title}
  {Nonreciprocal supercurrents in a field-free graphene {Josephson} triode},\
  }\href {https://doi.org/10.1021/acs.nanolett.3c01276} {\bibfield  {journal}
  {\bibinfo  {journal} {Nano Letters}\ }\textbf {\bibinfo {volume} {23}},\
  \bibinfo {pages} {5257} (\bibinfo {year} {2023})}\BibitemShut {NoStop}%
\bibitem [{\citenamefont {Zhang}\ \emph {et~al.}(2023)\citenamefont {Zhang},
  \citenamefont {Shao}, \citenamefont {Wang}, \citenamefont {Yang},
  \citenamefont {Yang},\ and\ \citenamefont
  {Tsymbal}}]{PhysRevLett.131.246301}%
  \BibitemOpen
  \bibfield  {author} {\bibinfo {author} {\bibfnamefont {S.-H.}\ \bibnamefont
  {Zhang}}, \bibinfo {author} {\bibfnamefont {D.-F.}\ \bibnamefont {Shao}},
  \bibinfo {author} {\bibfnamefont {Z.-A.}\ \bibnamefont {Wang}}, \bibinfo
  {author} {\bibfnamefont {J.}~\bibnamefont {Yang}}, \bibinfo {author}
  {\bibfnamefont {W.}~\bibnamefont {Yang}},\ and\ \bibinfo {author}
  {\bibfnamefont {E.~Y.}\ \bibnamefont {Tsymbal}},\ }\bibfield  {title}
  {\bibinfo {title} {Tunneling valley {Hall} effect driven by tilted {Dirac}
  fermions},\ }\href {https://doi.org/10.1103/PhysRevLett.131.246301}
  {\bibfield  {journal} {\bibinfo  {journal} {Phys. Rev. Lett.}\ }\textbf
  {\bibinfo {volume} {131}},\ \bibinfo {pages} {246301} (\bibinfo {year}
  {2023})}\BibitemShut {NoStop}%
\bibitem [{\citenamefont {Zeng}(2024)}]{PhysRevB.110.024511}%
  \BibitemOpen
  \bibfield  {author} {\bibinfo {author} {\bibfnamefont {W.}~\bibnamefont
  {Zeng}},\ }\bibfield  {title} {\bibinfo {title} {Tunneling chirality {Hall}
  effect in {type-I} {Weyl} semimetals},\ }\href
  {https://doi.org/10.1103/PhysRevB.110.024511} {\bibfield  {journal} {\bibinfo
   {journal} {Phys. Rev. B}\ }\textbf {\bibinfo {volume} {110}},\ \bibinfo
  {pages} {024511} (\bibinfo {year} {2024})}\BibitemShut {NoStop}%
\bibitem [{\citenamefont {Das}\ \emph {et~al.}(2024)\citenamefont {Das},
  \citenamefont {Ghorai}, \citenamefont {Culcer},\ and\ \citenamefont
  {Agarwal}}]{PhysRevLett.132.096302}%
  \BibitemOpen
  \bibfield  {author} {\bibinfo {author} {\bibfnamefont {K.}~\bibnamefont
  {Das}}, \bibinfo {author} {\bibfnamefont {K.}~\bibnamefont {Ghorai}},
  \bibinfo {author} {\bibfnamefont {D.}~\bibnamefont {Culcer}},\ and\ \bibinfo
  {author} {\bibfnamefont {A.}~\bibnamefont {Agarwal}},\ }\bibfield  {title}
  {\bibinfo {title} {Nonlinear valley {Hall} effect},\ }\href
  {https://doi.org/10.1103/PhysRevLett.132.096302} {\bibfield  {journal}
  {\bibinfo  {journal} {Phys. Rev. Lett.}\ }\textbf {\bibinfo {volume} {132}},\
  \bibinfo {pages} {096302} (\bibinfo {year} {2024})}\BibitemShut {NoStop}%
\bibitem [{\citenamefont {Avsar}\ \emph {et~al.}(2020)\citenamefont {Avsar},
  \citenamefont {Ochoa}, \citenamefont {Guinea}, \citenamefont {\"Ozyilmaz},
  \citenamefont {van Wees},\ and\ \citenamefont
  {Vera-Marun}}]{RevModPhys.92.021003}%
  \BibitemOpen
  \bibfield  {author} {\bibinfo {author} {\bibfnamefont {A.}~\bibnamefont
  {Avsar}}, \bibinfo {author} {\bibfnamefont {H.}~\bibnamefont {Ochoa}},
  \bibinfo {author} {\bibfnamefont {F.}~\bibnamefont {Guinea}}, \bibinfo
  {author} {\bibfnamefont {B.}~\bibnamefont {\"Ozyilmaz}}, \bibinfo {author}
  {\bibfnamefont {B.~J.}\ \bibnamefont {van Wees}},\ and\ \bibinfo {author}
  {\bibfnamefont {I.~J.}\ \bibnamefont {Vera-Marun}},\ }\bibfield  {title}
  {\bibinfo {title} {Colloquium: {Spintronics} in graphene and other
  two-dimensional materials},\ }\href
  {https://doi.org/10.1103/RevModPhys.92.021003} {\bibfield  {journal}
  {\bibinfo  {journal} {Rev. Mod. Phys.}\ }\textbf {\bibinfo {volume} {92}},\
  \bibinfo {pages} {021003} (\bibinfo {year} {2020})}\BibitemShut {NoStop}%
\bibitem [{\citenamefont {Armitage}\ \emph {et~al.}(2018)\citenamefont
  {Armitage}, \citenamefont {Mele},\ and\ \citenamefont
  {Vishwanath}}]{RevModPhys.90.015001}%
  \BibitemOpen
  \bibfield  {author} {\bibinfo {author} {\bibfnamefont {N.~P.}\ \bibnamefont
  {Armitage}}, \bibinfo {author} {\bibfnamefont {E.~J.}\ \bibnamefont {Mele}},\
  and\ \bibinfo {author} {\bibfnamefont {A.}~\bibnamefont {Vishwanath}},\
  }\bibfield  {title} {\bibinfo {title} {Weyl and {Dirac} semimetals in
  three-dimensional solids},\ }\href
  {https://doi.org/10.1103/RevModPhys.90.015001} {\bibfield  {journal}
  {\bibinfo  {journal} {Rev. Mod. Phys.}\ }\textbf {\bibinfo {volume} {90}},\
  \bibinfo {pages} {015001} (\bibinfo {year} {2018})}\BibitemShut {NoStop}%
\bibitem [{\citenamefont {Beenakker}(2006)}]{PhysRevLett.97.067007}%
  \BibitemOpen
  \bibfield  {author} {\bibinfo {author} {\bibfnamefont {C.~W.~J.}\
  \bibnamefont {Beenakker}},\ }\bibfield  {title} {\bibinfo {title} {Specular
  {Andreev} reflection in graphene},\ }\href
  {https://doi.org/10.1103/PhysRevLett.97.067007} {\bibfield  {journal}
  {\bibinfo  {journal} {Phys. Rev. Lett.}\ }\textbf {\bibinfo {volume} {97}},\
  \bibinfo {pages} {067007} (\bibinfo {year} {2006})}\BibitemShut {NoStop}%
\bibitem [{\citenamefont {Zeng}\ and\ \citenamefont
  {Shen}(2022{\natexlab{a}})}]{Zeng_2022}%
  \BibitemOpen
  \bibfield  {author} {\bibinfo {author} {\bibfnamefont {W.}~\bibnamefont
  {Zeng}}\ and\ \bibinfo {author} {\bibfnamefont {R.}~\bibnamefont {Shen}},\
  }\bibfield  {title} {\bibinfo {title} {Andreev reflection of massive
  pseudospin-1 fermions},\ }\href {https://doi.org/10.1088/1367-2630/ac614e}
  {\bibfield  {journal} {\bibinfo  {journal} {New Journal of Physics}\ }\textbf
  {\bibinfo {volume} {24}},\ \bibinfo {pages} {043021} (\bibinfo {year}
  {2022}{\natexlab{a}})}\BibitemShut {NoStop}%
\bibitem [{\citenamefont {Huang}\ \emph {et~al.}(2023)\citenamefont {Huang},
  \citenamefont {Zeng},\ and\ \citenamefont {Shen}}]{HUANG2023128671}%
  \BibitemOpen
  \bibfield  {author} {\bibinfo {author} {\bibfnamefont {Y.}~\bibnamefont
  {Huang}}, \bibinfo {author} {\bibfnamefont {W.}~\bibnamefont {Zeng}},\ and\
  \bibinfo {author} {\bibfnamefont {R.}~\bibnamefont {Shen}},\ }\bibfield
  {title} {\bibinfo {title} {Photoinduced {Klein} tunneling in {semi-Dirac}
  material},\ }\href
  {https://doi.org/https://doi.org/10.1016/j.physleta.2023.128671} {\bibfield
  {journal} {\bibinfo  {journal} {Physics Letters A}\ }\textbf {\bibinfo
  {volume} {463}},\ \bibinfo {pages} {128671} (\bibinfo {year}
  {2023})}\BibitemShut {NoStop}%
\bibitem [{\citenamefont {Jafari}(2019)}]{PhysRevB.100.045144}%
  \BibitemOpen
  \bibfield  {author} {\bibinfo {author} {\bibfnamefont {S.~A.}\ \bibnamefont
  {Jafari}},\ }\bibfield  {title} {\bibinfo {title} {Electric field assisted
  amplification of magnetic fields in tilted {Dirac} cone systems},\ }\href
  {https://doi.org/10.1103/PhysRevB.100.045144} {\bibfield  {journal} {\bibinfo
   {journal} {Phys. Rev. B}\ }\textbf {\bibinfo {volume} {100}},\ \bibinfo
  {pages} {045144} (\bibinfo {year} {2019})}\BibitemShut {NoStop}%
\bibitem [{\citenamefont {Goerbig}\ \emph {et~al.}(2008)\citenamefont
  {Goerbig}, \citenamefont {Fuchs}, \citenamefont {Montambaux},\ and\
  \citenamefont {Pi\'echon}}]{PhysRevB.78.045415}%
  \BibitemOpen
  \bibfield  {author} {\bibinfo {author} {\bibfnamefont {M.~O.}\ \bibnamefont
  {Goerbig}}, \bibinfo {author} {\bibfnamefont {J.-N.}\ \bibnamefont {Fuchs}},
  \bibinfo {author} {\bibfnamefont {G.}~\bibnamefont {Montambaux}},\ and\
  \bibinfo {author} {\bibfnamefont {F.}~\bibnamefont {Pi\'echon}},\ }\bibfield
  {title} {\bibinfo {title} {Tilted anisotropic {Dirac} cones in quinoid-type
  graphene and
  {$\ensuremath{\alpha}\text{\ensuremath{-}}{(\text{BEDT-TTF})}_{2}{\text{I}}_{3}$}},\
  }\href {https://doi.org/10.1103/PhysRevB.78.045415} {\bibfield  {journal}
  {\bibinfo  {journal} {Phys. Rev. B}\ }\textbf {\bibinfo {volume} {78}},\
  \bibinfo {pages} {045415} (\bibinfo {year} {2008})}\BibitemShut {NoStop}%
\bibitem [{\citenamefont {Zhou}\ \emph {et~al.}(2014)\citenamefont {Zhou},
  \citenamefont {Dong}, \citenamefont {Oganov}, \citenamefont {Zhu},
  \citenamefont {Tian},\ and\ \citenamefont {Wang}}]{PhysRevLett.112.085502}%
  \BibitemOpen
  \bibfield  {author} {\bibinfo {author} {\bibfnamefont {X.-F.}\ \bibnamefont
  {Zhou}}, \bibinfo {author} {\bibfnamefont {X.}~\bibnamefont {Dong}}, \bibinfo
  {author} {\bibfnamefont {A.~R.}\ \bibnamefont {Oganov}}, \bibinfo {author}
  {\bibfnamefont {Q.}~\bibnamefont {Zhu}}, \bibinfo {author} {\bibfnamefont
  {Y.}~\bibnamefont {Tian}},\ and\ \bibinfo {author} {\bibfnamefont {H.-T.}\
  \bibnamefont {Wang}},\ }\bibfield  {title} {\bibinfo {title} {Semimetallic
  two-dimensional boron allotrope with massless {Dirac} fermions},\ }\href
  {https://doi.org/10.1103/PhysRevLett.112.085502} {\bibfield  {journal}
  {\bibinfo  {journal} {Phys. Rev. Lett.}\ }\textbf {\bibinfo {volume} {112}},\
  \bibinfo {pages} {085502} (\bibinfo {year} {2014})}\BibitemShut {NoStop}%
\bibitem [{\citenamefont {Lopez-Bezanilla}\ and\ \citenamefont
  {Littlewood}(2016)}]{PhysRevB.93.241405}%
  \BibitemOpen
  \bibfield  {author} {\bibinfo {author} {\bibfnamefont {A.}~\bibnamefont
  {Lopez-Bezanilla}}\ and\ \bibinfo {author} {\bibfnamefont {P.~B.}\
  \bibnamefont {Littlewood}},\ }\bibfield  {title} {\bibinfo {title}
  {Electronic properties of $8\text{\ensuremath{-}}\mathit{Pmmn}$ borophene},\
  }\href {https://doi.org/10.1103/PhysRevB.93.241405} {\bibfield  {journal}
  {\bibinfo  {journal} {Phys. Rev. B}\ }\textbf {\bibinfo {volume} {93}},\
  \bibinfo {pages} {241405} (\bibinfo {year} {2016})}\BibitemShut {NoStop}%
\bibitem [{\citenamefont {Nakhaee}\ \emph {et~al.}(2018)\citenamefont
  {Nakhaee}, \citenamefont {Ketabi},\ and\ \citenamefont
  {Peeters}}]{PhysRevB.97.125424}%
  \BibitemOpen
  \bibfield  {author} {\bibinfo {author} {\bibfnamefont {M.}~\bibnamefont
  {Nakhaee}}, \bibinfo {author} {\bibfnamefont {S.~A.}\ \bibnamefont
  {Ketabi}},\ and\ \bibinfo {author} {\bibfnamefont {F.~M.}\ \bibnamefont
  {Peeters}},\ }\bibfield  {title} {\bibinfo {title} {Tight-binding model for
  borophene and borophane},\ }\href
  {https://doi.org/10.1103/PhysRevB.97.125424} {\bibfield  {journal} {\bibinfo
  {journal} {Phys. Rev. B}\ }\textbf {\bibinfo {volume} {97}},\ \bibinfo
  {pages} {125424} (\bibinfo {year} {2018})}\BibitemShut {NoStop}%
\bibitem [{\citenamefont {Katayama}\ \emph {et~al.}(2006)\citenamefont
  {Katayama}, \citenamefont {Kobayashi},\ and\ \citenamefont
  {Suzumura}}]{doi:10.1143/JPSJ.75.054705}%
  \BibitemOpen
  \bibfield  {author} {\bibinfo {author} {\bibfnamefont {S.}~\bibnamefont
  {Katayama}}, \bibinfo {author} {\bibfnamefont {A.}~\bibnamefont
  {Kobayashi}},\ and\ \bibinfo {author} {\bibfnamefont {Y.}~\bibnamefont
  {Suzumura}},\ }\bibfield  {title} {\bibinfo {title} {Pressure-induced
  zero-gap semiconducting state in organic conductor
  {$\ensuremath{\alpha}\text{\ensuremath{-}}{(\text{BEDT-TTF})}_{2}{\text{I}}_{3}$}
  salt},\ }\href {https://doi.org/10.1143/JPSJ.75.054705} {\bibfield  {journal}
  {\bibinfo  {journal} {Journal of the Physical Society of Japan}\ }\textbf
  {\bibinfo {volume} {75}},\ \bibinfo {pages} {054705} (\bibinfo {year}
  {2006})}\BibitemShut {NoStop}%
\bibitem [{\citenamefont {Tan}\ \emph {et~al.}(2021)\citenamefont {Tan},
  \citenamefont {Yan}, \citenamefont {Zhao}, \citenamefont {Guo},\ and\
  \citenamefont {Chang}}]{PhysRevB.103.125425}%
  \BibitemOpen
  \bibfield  {author} {\bibinfo {author} {\bibfnamefont {C.-Y.}\ \bibnamefont
  {Tan}}, \bibinfo {author} {\bibfnamefont {C.-X.}\ \bibnamefont {Yan}},
  \bibinfo {author} {\bibfnamefont {Y.-H.}\ \bibnamefont {Zhao}}, \bibinfo
  {author} {\bibfnamefont {H.}~\bibnamefont {Guo}},\ and\ \bibinfo {author}
  {\bibfnamefont {H.-R.}\ \bibnamefont {Chang}},\ }\bibfield  {title} {\bibinfo
  {title} {Anisotropic longitudinal optical conductivities of tilted {Dirac}
  bands in
  {$1{T}^{\ensuremath{'}}\text{\ensuremath{-}}\mathrm{Mo}{\mathrm{S}}_{2}$}},\
  }\href {https://doi.org/10.1103/PhysRevB.103.125425} {\bibfield  {journal}
  {\bibinfo  {journal} {Phys. Rev. B}\ }\textbf {\bibinfo {volume} {103}},\
  \bibinfo {pages} {125425} (\bibinfo {year} {2021})}\BibitemShut {NoStop}%
\bibitem [{\citenamefont {Qian}\ \emph {et~al.}(2014)\citenamefont {Qian},
  \citenamefont {Liu}, \citenamefont {Fu},\ and\ \citenamefont {Li}}]{sxss5}%
  \BibitemOpen
  \bibfield  {author} {\bibinfo {author} {\bibfnamefont {X.}~\bibnamefont
  {Qian}}, \bibinfo {author} {\bibfnamefont {J.}~\bibnamefont {Liu}}, \bibinfo
  {author} {\bibfnamefont {L.}~\bibnamefont {Fu}},\ and\ \bibinfo {author}
  {\bibfnamefont {J.}~\bibnamefont {Li}},\ }\bibfield  {title} {\bibinfo
  {title} {Quantum spin {Hall} effect in two-dimensional transition metal
  dichalcogenides},\ }\href {https://doi.org/10.1126/science.1256815}
  {\bibfield  {journal} {\bibinfo  {journal} {Science}\ }\textbf {\bibinfo
  {volume} {346}},\ \bibinfo {pages} {1344} (\bibinfo {year}
  {2014})}\BibitemShut {NoStop}%
\bibitem [{\citenamefont {Wang}\ \emph {et~al.}(2019)\citenamefont {Wang},
  \citenamefont {L{\"u}}, \citenamefont {Wang}, \citenamefont {Feng},\ and\
  \citenamefont {Zheng}}]{doi:10.1021/acsaelm.9b00017}%
  \BibitemOpen
  \bibfield  {author} {\bibinfo {author} {\bibfnamefont {Z.-Q.}\ \bibnamefont
  {Wang}}, \bibinfo {author} {\bibfnamefont {T.-Y.}\ \bibnamefont {L{\"u}}},
  \bibinfo {author} {\bibfnamefont {H.-Q.}\ \bibnamefont {Wang}}, \bibinfo
  {author} {\bibfnamefont {Y.~P.}\ \bibnamefont {Feng}},\ and\ \bibinfo
  {author} {\bibfnamefont {J.-C.}\ \bibnamefont {Zheng}},\ }\bibfield  {title}
  {\bibinfo {title} {Band gap opening in {8-Pmmn} borophene by hydrogenation},\
  }\href {https://doi.org/10.1021/acsaelm.9b00017} {\bibfield  {journal}
  {\bibinfo  {journal} {ACS Applied Electronic Materials}\ }\textbf {\bibinfo
  {volume} {1}},\ \bibinfo {pages} {667} (\bibinfo {year} {2019})}\BibitemShut
  {NoStop}%
\bibitem [{\citenamefont {Mojarro}\ \emph {et~al.}(2021)\citenamefont
  {Mojarro}, \citenamefont {Carrillo-Bastos},\ and\ \citenamefont
  {Maytorena}}]{PhysRevB.103.165415}%
  \BibitemOpen
  \bibfield  {author} {\bibinfo {author} {\bibfnamefont {M.~A.}\ \bibnamefont
  {Mojarro}}, \bibinfo {author} {\bibfnamefont {R.}~\bibnamefont
  {Carrillo-Bastos}},\ and\ \bibinfo {author} {\bibfnamefont {J.~A.}\
  \bibnamefont {Maytorena}},\ }\bibfield  {title} {\bibinfo {title} {Optical
  properties of massive anisotropic tilted {Dirac} systems},\ }\href
  {https://doi.org/10.1103/PhysRevB.103.165415} {\bibfield  {journal} {\bibinfo
   {journal} {Phys. Rev. B}\ }\textbf {\bibinfo {volume} {103}},\ \bibinfo
  {pages} {165415} (\bibinfo {year} {2021})}\BibitemShut {NoStop}%
\bibitem [{\citenamefont {Kapri}\ \emph {et~al.}(2020)\citenamefont {Kapri},
  \citenamefont {Dey},\ and\ \citenamefont {Ghosh}}]{PhysRevB.102.045417}%
  \BibitemOpen
  \bibfield  {author} {\bibinfo {author} {\bibfnamefont {P.}~\bibnamefont
  {Kapri}}, \bibinfo {author} {\bibfnamefont {B.}~\bibnamefont {Dey}},\ and\
  \bibinfo {author} {\bibfnamefont {T.~K.}\ \bibnamefont {Ghosh}},\ }\bibfield
  {title} {\bibinfo {title} {Valley caloritronics in a photodriven
  heterojunction of {Dirac} materials},\ }\href
  {https://doi.org/10.1103/PhysRevB.102.045417} {\bibfield  {journal} {\bibinfo
   {journal} {Phys. Rev. B}\ }\textbf {\bibinfo {volume} {102}},\ \bibinfo
  {pages} {045417} (\bibinfo {year} {2020})}\BibitemShut {NoStop}%
\bibitem [{\citenamefont {Furusaki}\ and\ \citenamefont
  {Tsukada}(1991)}]{FURUSAKI1991299}%
  \BibitemOpen
  \bibfield  {author} {\bibinfo {author} {\bibfnamefont {A.}~\bibnamefont
  {Furusaki}}\ and\ \bibinfo {author} {\bibfnamefont {M.}~\bibnamefont
  {Tsukada}},\ }\bibfield  {title} {\bibinfo {title} {Dc {Josephson} effect and
  {Andreev} reflection},\ }\href
  {https://doi.org/https://doi.org/10.1016/0038-1098(91)90201-6} {\bibfield
  {journal} {\bibinfo  {journal} {Solid State Communications}\ }\textbf
  {\bibinfo {volume} {78}},\ \bibinfo {pages} {299} (\bibinfo {year}
  {1991})}\BibitemShut {NoStop}%
\bibitem [{\citenamefont {Zhao}\ and\ \citenamefont
  {Shen}(2006)}]{PhysRevB.73.214511}%
  \BibitemOpen
  \bibfield  {author} {\bibinfo {author} {\bibfnamefont {Y.}~\bibnamefont
  {Zhao}}\ and\ \bibinfo {author} {\bibfnamefont {R.}~\bibnamefont {Shen}},\
  }\bibfield  {title} {\bibinfo {title} {Josephson effect in junctions of
  ferromagnetic superconductors with equal spin pairing symmetry},\ }\href
  {https://doi.org/10.1103/PhysRevB.73.214511} {\bibfield  {journal} {\bibinfo
  {journal} {Phys. Rev. B}\ }\textbf {\bibinfo {volume} {73}},\ \bibinfo
  {pages} {214511} (\bibinfo {year} {2006})}\BibitemShut {NoStop}%
\bibitem [{\citenamefont {Lu}\ \emph {et~al.}(2024)\citenamefont {Lu},
  \citenamefont {Maeda}, \citenamefont {Ito}, \citenamefont {Yada},\ and\
  \citenamefont {Tanaka}}]{PhysRevLett.133.226002}%
  \BibitemOpen
  \bibfield  {author} {\bibinfo {author} {\bibfnamefont {B.}~\bibnamefont
  {Lu}}, \bibinfo {author} {\bibfnamefont {K.}~\bibnamefont {Maeda}}, \bibinfo
  {author} {\bibfnamefont {H.}~\bibnamefont {Ito}}, \bibinfo {author}
  {\bibfnamefont {K.}~\bibnamefont {Yada}},\ and\ \bibinfo {author}
  {\bibfnamefont {Y.}~\bibnamefont {Tanaka}},\ }\bibfield  {title} {\bibinfo
  {title} {$\ensuremath{\varphi}$ {Josephson} junction induced by
  altermagnetism},\ }\href {https://doi.org/10.1103/PhysRevLett.133.226002}
  {\bibfield  {journal} {\bibinfo  {journal} {Phys. Rev. Lett.}\ }\textbf
  {\bibinfo {volume} {133}},\ \bibinfo {pages} {226002} (\bibinfo {year}
  {2024})}\BibitemShut {NoStop}%
\bibitem [{\citenamefont {McMillan}(1968)}]{PhysRev.175.559}%
  \BibitemOpen
  \bibfield  {author} {\bibinfo {author} {\bibfnamefont {W.~L.}\ \bibnamefont
  {McMillan}},\ }\bibfield  {title} {\bibinfo {title} {Theory of
  superconductor---normal-metal interfaces},\ }\href
  {https://doi.org/10.1103/PhysRev.175.559} {\bibfield  {journal} {\bibinfo
  {journal} {Phys. Rev.}\ }\textbf {\bibinfo {volume} {175}},\ \bibinfo {pages}
  {559} (\bibinfo {year} {1968})}\BibitemShut {NoStop}%
\bibitem [{\citenamefont {Tanaka}\ and\ \citenamefont
  {Kashiwaya}(1997)}]{PhysRevB.56.892}%
  \BibitemOpen
  \bibfield  {author} {\bibinfo {author} {\bibfnamefont {Y.}~\bibnamefont
  {Tanaka}}\ and\ \bibinfo {author} {\bibfnamefont {S.}~\bibnamefont
  {Kashiwaya}},\ }\bibfield  {title} {\bibinfo {title} {Theory of {Josephson}
  effects in anisotropic superconductors},\ }\href
  {https://doi.org/10.1103/PhysRevB.56.892} {\bibfield  {journal} {\bibinfo
  {journal} {Phys. Rev. B}\ }\textbf {\bibinfo {volume} {56}},\ \bibinfo
  {pages} {892} (\bibinfo {year} {1997})}\BibitemShut {NoStop}%
\bibitem [{\citenamefont {Mojarro}\ \emph {et~al.}(2022)\citenamefont
  {Mojarro}, \citenamefont {Carrillo-Bastos},\ and\ \citenamefont
  {Maytorena}}]{PhysRevB.105.L201408}%
  \BibitemOpen
  \bibfield  {author} {\bibinfo {author} {\bibfnamefont {M.~A.}\ \bibnamefont
  {Mojarro}}, \bibinfo {author} {\bibfnamefont {R.}~\bibnamefont
  {Carrillo-Bastos}},\ and\ \bibinfo {author} {\bibfnamefont {J.~A.}\
  \bibnamefont {Maytorena}},\ }\bibfield  {title} {\bibinfo {title} {Hyperbolic
  plasmons in massive tilted two-dimensional {Dirac} materials},\ }\href
  {https://doi.org/10.1103/PhysRevB.105.L201408} {\bibfield  {journal}
  {\bibinfo  {journal} {Phys. Rev. B}\ }\textbf {\bibinfo {volume} {105}},\
  \bibinfo {pages} {L201408} (\bibinfo {year} {2022})}\BibitemShut {NoStop}%
\bibitem [{\citenamefont {Tan}\ \emph {et~al.}(2022)\citenamefont {Tan},
  \citenamefont {Hou}, \citenamefont {Yan}, \citenamefont {Guo},\ and\
  \citenamefont {Chang}}]{PhysRevB.106.165404}%
  \BibitemOpen
  \bibfield  {author} {\bibinfo {author} {\bibfnamefont {C.-Y.}\ \bibnamefont
  {Tan}}, \bibinfo {author} {\bibfnamefont {J.-T.}\ \bibnamefont {Hou}},
  \bibinfo {author} {\bibfnamefont {C.-X.}\ \bibnamefont {Yan}}, \bibinfo
  {author} {\bibfnamefont {H.}~\bibnamefont {Guo}},\ and\ \bibinfo {author}
  {\bibfnamefont {H.-R.}\ \bibnamefont {Chang}},\ }\bibfield  {title} {\bibinfo
  {title} {Signatures of {Lifshitz} transition in the optical conductivity of
  two-dimensional tilted {Dirac} materials},\ }\href
  {https://doi.org/10.1103/PhysRevB.106.165404} {\bibfield  {journal} {\bibinfo
   {journal} {Phys. Rev. B}\ }\textbf {\bibinfo {volume} {106}},\ \bibinfo
  {pages} {165404} (\bibinfo {year} {2022})}\BibitemShut {NoStop}%
\bibitem [{\citenamefont {Ben~Shalom}\ \emph {et~al.}(2016)\citenamefont
  {Ben~Shalom}, \citenamefont {Zhu}, \citenamefont {Fal’Ko}, \citenamefont
  {Mishchenko}, \citenamefont {Kretinin}, \citenamefont {Novoselov},
  \citenamefont {Woods}, \citenamefont {Watanabe}, \citenamefont {Taniguchi},
  \citenamefont {Geim} \emph {et~al.}}]{ben2016quantum}%
  \BibitemOpen
  \bibfield  {author} {\bibinfo {author} {\bibfnamefont {M.}~\bibnamefont
  {Ben~Shalom}}, \bibinfo {author} {\bibfnamefont {M.}~\bibnamefont {Zhu}},
  \bibinfo {author} {\bibfnamefont {V.}~\bibnamefont {Fal’Ko}}, \bibinfo
  {author} {\bibfnamefont {A.}~\bibnamefont {Mishchenko}}, \bibinfo {author}
  {\bibfnamefont {A.}~\bibnamefont {Kretinin}}, \bibinfo {author}
  {\bibfnamefont {K.}~\bibnamefont {Novoselov}}, \bibinfo {author}
  {\bibfnamefont {C.}~\bibnamefont {Woods}}, \bibinfo {author} {\bibfnamefont
  {K.}~\bibnamefont {Watanabe}}, \bibinfo {author} {\bibfnamefont
  {T.}~\bibnamefont {Taniguchi}}, \bibinfo {author} {\bibfnamefont
  {A.}~\bibnamefont {Geim}}, \emph {et~al.},\ }\bibfield  {title} {\bibinfo
  {title} {Quantum oscillations of the critical current and high-field
  superconducting proximity in ballistic graphene},\ }\href
  {https://doi.org/10.1038/nphys3592} {\bibfield  {journal} {\bibinfo
  {journal} {Nature Physics}\ }\textbf {\bibinfo {volume} {12}},\ \bibinfo
  {pages} {318} (\bibinfo {year} {2016})}\BibitemShut {NoStop}%
\bibitem [{\citenamefont {Burset}\ \emph {et~al.}(2008)\citenamefont {Burset},
  \citenamefont {Yeyati},\ and\ \citenamefont
  {Mart\'{\i}n-Rodero}}]{PhysRevB.77.205425}%
  \BibitemOpen
  \bibfield  {author} {\bibinfo {author} {\bibfnamefont {P.}~\bibnamefont
  {Burset}}, \bibinfo {author} {\bibfnamefont {A.~L.}\ \bibnamefont {Yeyati}},\
  and\ \bibinfo {author} {\bibfnamefont {A.}~\bibnamefont
  {Mart\'{\i}n-Rodero}},\ }\bibfield  {title} {\bibinfo {title} {Microscopic
  theory of the proximity effect in superconductor-graphene nanostructures},\
  }\href {https://doi.org/10.1103/PhysRevB.77.205425} {\bibfield  {journal}
  {\bibinfo  {journal} {Phys. Rev. B}\ }\textbf {\bibinfo {volume} {77}},\
  \bibinfo {pages} {205425} (\bibinfo {year} {2008})}\BibitemShut {NoStop}%
\bibitem [{\citenamefont {Coskun}\ \emph {et~al.}(2012)\citenamefont {Coskun},
  \citenamefont {Brenner}, \citenamefont {Hymel}, \citenamefont {Vakaryuk},
  \citenamefont {Levchenko},\ and\ \citenamefont
  {Bezryadin}}]{PhysRevLett.108.097003}%
  \BibitemOpen
  \bibfield  {author} {\bibinfo {author} {\bibfnamefont {U.~C.}\ \bibnamefont
  {Coskun}}, \bibinfo {author} {\bibfnamefont {M.}~\bibnamefont {Brenner}},
  \bibinfo {author} {\bibfnamefont {T.}~\bibnamefont {Hymel}}, \bibinfo
  {author} {\bibfnamefont {V.}~\bibnamefont {Vakaryuk}}, \bibinfo {author}
  {\bibfnamefont {A.}~\bibnamefont {Levchenko}},\ and\ \bibinfo {author}
  {\bibfnamefont {A.}~\bibnamefont {Bezryadin}},\ }\bibfield  {title} {\bibinfo
  {title} {Distribution of supercurrent switching in graphene under the
  proximity effect},\ }\href {https://doi.org/10.1103/PhysRevLett.108.097003}
  {\bibfield  {journal} {\bibinfo  {journal} {Phys. Rev. Lett.}\ }\textbf
  {\bibinfo {volume} {108}},\ \bibinfo {pages} {097003} (\bibinfo {year}
  {2012})}\BibitemShut {NoStop}%
\bibitem [{\citenamefont {Beenakker}(2008)}]{RevModPhys.80.1337}%
  \BibitemOpen
  \bibfield  {author} {\bibinfo {author} {\bibfnamefont {C.~W.~J.}\
  \bibnamefont {Beenakker}},\ }\bibfield  {title} {\bibinfo {title}
  {Colloquium: {Andreev} reflection and {Klein} tunneling in graphene},\ }\href
  {https://doi.org/10.1103/RevModPhys.80.1337} {\bibfield  {journal} {\bibinfo
  {journal} {Rev. Mod. Phys.}\ }\textbf {\bibinfo {volume} {80}},\ \bibinfo
  {pages} {1337} (\bibinfo {year} {2008})}\BibitemShut {NoStop}%
\bibitem [{\citenamefont {De~Gennes}(2018)}]{de2018superconductivity}%
  \BibitemOpen
  \bibfield  {author} {\bibinfo {author} {\bibfnamefont {P.-G.}\ \bibnamefont
  {De~Gennes}},\ }\href@noop {} {\emph {\bibinfo {title} {Superconductivity of
  metals and alloys}}}\ (\bibinfo  {publisher} {CRC press},\ \bibinfo {year}
  {2018})\BibitemShut {NoStop}%
\bibitem [{\citenamefont {Brydon}\ \emph {et~al.}(2013)\citenamefont {Brydon},
  \citenamefont {Timm},\ and\ \citenamefont {Schnyder}}]{Brydon_2013}%
  \BibitemOpen
  \bibfield  {author} {\bibinfo {author} {\bibfnamefont {P.~M.~R.}\
  \bibnamefont {Brydon}}, \bibinfo {author} {\bibfnamefont {C.}~\bibnamefont
  {Timm}},\ and\ \bibinfo {author} {\bibfnamefont {A.~P.}\ \bibnamefont
  {Schnyder}},\ }\bibfield  {title} {\bibinfo {title} {Interface currents in
  topological superconductor-ferromagnet heterostructures},\ }\href
  {https://doi.org/10.1088/1367-2630/15/4/045019} {\bibfield  {journal}
  {\bibinfo  {journal} {New Journal of Physics}\ }\textbf {\bibinfo {volume}
  {15}},\ \bibinfo {pages} {045019} (\bibinfo {year} {2013})}\BibitemShut
  {NoStop}%
\bibitem [{\citenamefont {Paul}\ \emph {et~al.}(2019)\citenamefont {Paul},
  \citenamefont {Islam},\ and\ \citenamefont {Saha}}]{PhysRevB.99.155418}%
  \BibitemOpen
  \bibfield  {author} {\bibinfo {author} {\bibfnamefont {G.~C.}\ \bibnamefont
  {Paul}}, \bibinfo {author} {\bibfnamefont {S.~F.}\ \bibnamefont {Islam}},\
  and\ \bibinfo {author} {\bibfnamefont {A.}~\bibnamefont {Saha}},\ }\bibfield
  {title} {\bibinfo {title} {Fingerprints of tilted {Dirac} cones on the {RKKY}
  exchange interaction in $8\text{\ensuremath{-}}\mathit{Pmmn}$ borophene},\
  }\href {https://doi.org/10.1103/PhysRevB.99.155418} {\bibfield  {journal}
  {\bibinfo  {journal} {Phys. Rev. B}\ }\textbf {\bibinfo {volume} {99}},\
  \bibinfo {pages} {155418} (\bibinfo {year} {2019})}\BibitemShut {NoStop}%
\bibitem [{\citenamefont {Vanderstraeten}\ and\ \citenamefont
  {Vande~Ginste}(2024)}]{PhysRevB.109.205413}%
  \BibitemOpen
  \bibfield  {author} {\bibinfo {author} {\bibfnamefont {E.}~\bibnamefont
  {Vanderstraeten}}\ and\ \bibinfo {author} {\bibfnamefont {D.}~\bibnamefont
  {Vande~Ginste}},\ }\bibfield  {title} {\bibinfo {title} {Valley filtering in
  $8\text{\ensuremath{-}}\mathit{Pmmn}$ borophene based on an electrostatic
  waveguide constriction},\ }\href
  {https://doi.org/10.1103/PhysRevB.109.205413} {\bibfield  {journal} {\bibinfo
   {journal} {Phys. Rev. B}\ }\textbf {\bibinfo {volume} {109}},\ \bibinfo
  {pages} {205413} (\bibinfo {year} {2024})}\BibitemShut {NoStop}%
\bibitem [{\citenamefont {Champo}\ and\ \citenamefont
  {Naumis}(2019)}]{PhysRevB.99.035415}%
  \BibitemOpen
  \bibfield  {author} {\bibinfo {author} {\bibfnamefont {A.~E.}\ \bibnamefont
  {Champo}}\ and\ \bibinfo {author} {\bibfnamefont {G.~G.}\ \bibnamefont
  {Naumis}},\ }\bibfield  {title} {\bibinfo {title} {Metal-insulator transition
  in $8\ensuremath{-}pmmn$ borophene under normal incidence of electromagnetic
  radiation},\ }\href {https://doi.org/10.1103/PhysRevB.99.035415} {\bibfield
  {journal} {\bibinfo  {journal} {Phys. Rev. B}\ }\textbf {\bibinfo {volume}
  {99}},\ \bibinfo {pages} {035415} (\bibinfo {year} {2019})}\BibitemShut
  {NoStop}%
\bibitem [{\citenamefont {Buzdin}\ \emph {et~al.}(2011)\citenamefont {Buzdin},
  \citenamefont {Mel'nikov},\ and\ \citenamefont
  {Pugach}}]{PhysRevB.83.144515}%
  \BibitemOpen
  \bibfield  {author} {\bibinfo {author} {\bibfnamefont {A.~I.}\ \bibnamefont
  {Buzdin}}, \bibinfo {author} {\bibfnamefont {A.~S.}\ \bibnamefont
  {Mel'nikov}},\ and\ \bibinfo {author} {\bibfnamefont {N.~G.}\ \bibnamefont
  {Pugach}},\ }\bibfield  {title} {\bibinfo {title} {Domain walls and
  long-range triplet correlations in {SFS} {Josephson} junctions},\ }\href
  {https://doi.org/10.1103/PhysRevB.83.144515} {\bibfield  {journal} {\bibinfo
  {journal} {Phys. Rev. B}\ }\textbf {\bibinfo {volume} {83}},\ \bibinfo
  {pages} {144515} (\bibinfo {year} {2011})}\BibitemShut {NoStop}%
\bibitem [{\citenamefont {Crouzy}\ \emph {et~al.}(2007)\citenamefont {Crouzy},
  \citenamefont {Tollis},\ and\ \citenamefont {Ivanov}}]{PhysRevB.76.134502}%
  \BibitemOpen
  \bibfield  {author} {\bibinfo {author} {\bibfnamefont {B.}~\bibnamefont
  {Crouzy}}, \bibinfo {author} {\bibfnamefont {S.}~\bibnamefont {Tollis}},\
  and\ \bibinfo {author} {\bibfnamefont {D.~A.}\ \bibnamefont {Ivanov}},\
  }\bibfield  {title} {\bibinfo {title} {Josephson current in a
  superconductor-ferromagnet-superconductor junction with in-plane
  ferromagnetic domains},\ }\href {https://doi.org/10.1103/PhysRevB.76.134502}
  {\bibfield  {journal} {\bibinfo  {journal} {Phys. Rev. B}\ }\textbf {\bibinfo
  {volume} {76}},\ \bibinfo {pages} {134502} (\bibinfo {year}
  {2007})}\BibitemShut {NoStop}%
\bibitem [{\citenamefont {Wang}\ \emph {et~al.}(2014)\citenamefont {Wang},
  \citenamefont {Yang},\ and\ \citenamefont {Chan}}]{PhysRevB.89.064501}%
  \BibitemOpen
  \bibfield  {author} {\bibinfo {author} {\bibfnamefont {J.}~\bibnamefont
  {Wang}}, \bibinfo {author} {\bibfnamefont {Y.~H.}\ \bibnamefont {Yang}},\
  and\ \bibinfo {author} {\bibfnamefont {K.~S.}\ \bibnamefont {Chan}},\
  }\bibfield  {title} {\bibinfo {title} {Josephson $\ensuremath{\pi}$ state
  induced by valley polarization},\ }\href
  {https://doi.org/10.1103/PhysRevB.89.064501} {\bibfield  {journal} {\bibinfo
  {journal} {Phys. Rev. B}\ }\textbf {\bibinfo {volume} {89}},\ \bibinfo
  {pages} {064501} (\bibinfo {year} {2014})}\BibitemShut {NoStop}%
\bibitem [{\citenamefont {Matos-Abiague}\ and\ \citenamefont
  {Fabian}(2015)}]{PhysRevLett.115.056602}%
  \BibitemOpen
  \bibfield  {author} {\bibinfo {author} {\bibfnamefont {A.}~\bibnamefont
  {Matos-Abiague}}\ and\ \bibinfo {author} {\bibfnamefont {J.}~\bibnamefont
  {Fabian}},\ }\bibfield  {title} {\bibinfo {title} {Tunneling anomalous and
  spin {Hall} effects},\ }\href
  {https://doi.org/10.1103/PhysRevLett.115.056602} {\bibfield  {journal}
  {\bibinfo  {journal} {Phys. Rev. Lett.}\ }\textbf {\bibinfo {volume} {115}},\
  \bibinfo {pages} {056602} (\bibinfo {year} {2015})}\BibitemShut {NoStop}%
\bibitem [{\citenamefont {Scharf}\ \emph {et~al.}(2016)\citenamefont {Scharf},
  \citenamefont {Matos-Abiague}, \citenamefont {Han}, \citenamefont
  {Hankiewicz},\ and\ \citenamefont {\ifmmode \check{Z}\else
  \v{Z}\fi{}uti\ifmmode~\acute{c}\else \'{c}\fi{}}}]{PhysRevLett.117.166806}%
  \BibitemOpen
  \bibfield  {author} {\bibinfo {author} {\bibfnamefont {B.}~\bibnamefont
  {Scharf}}, \bibinfo {author} {\bibfnamefont {A.}~\bibnamefont
  {Matos-Abiague}}, \bibinfo {author} {\bibfnamefont {J.~E.}\ \bibnamefont
  {Han}}, \bibinfo {author} {\bibfnamefont {E.~M.}\ \bibnamefont
  {Hankiewicz}},\ and\ \bibinfo {author} {\bibfnamefont {I.}~\bibnamefont
  {\ifmmode \check{Z}\else \v{Z}\fi{}uti\ifmmode~\acute{c}\else \'{c}\fi{}}},\
  }\bibfield  {title} {\bibinfo {title} {Tunneling planar {Hall} effect in
  topological insulators: Spin valves and amplifiers},\ }\href
  {https://doi.org/10.1103/PhysRevLett.117.166806} {\bibfield  {journal}
  {\bibinfo  {journal} {Phys. Rev. Lett.}\ }\textbf {\bibinfo {volume} {117}},\
  \bibinfo {pages} {166806} (\bibinfo {year} {2016})}\BibitemShut {NoStop}%
\bibitem [{\citenamefont {Titov}\ and\ \citenamefont
  {Beenakker}(2006)}]{PhysRevB.74.041401}%
  \BibitemOpen
  \bibfield  {author} {\bibinfo {author} {\bibfnamefont {M.}~\bibnamefont
  {Titov}}\ and\ \bibinfo {author} {\bibfnamefont {C.~W.~J.}\ \bibnamefont
  {Beenakker}},\ }\bibfield  {title} {\bibinfo {title} {Josephson effect in
  ballistic graphene},\ }\href {https://doi.org/10.1103/PhysRevB.74.041401}
  {\bibfield  {journal} {\bibinfo  {journal} {Phys. Rev. B}\ }\textbf {\bibinfo
  {volume} {74}},\ \bibinfo {pages} {041401} (\bibinfo {year}
  {2006})}\BibitemShut {NoStop}%
\bibitem [{\citenamefont {Zeng}\ and\ \citenamefont
  {Shen}(2022{\natexlab{b}})}]{PhysRevB.105.094510}%
  \BibitemOpen
  \bibfield  {author} {\bibinfo {author} {\bibfnamefont {W.}~\bibnamefont
  {Zeng}}\ and\ \bibinfo {author} {\bibfnamefont {R.}~\bibnamefont {Shen}},\
  }\bibfield  {title} {\bibinfo {title} {Light-modulated {Josephson} effect in
  {Kekul\'e} patterned graphene},\ }\href
  {https://doi.org/10.1103/PhysRevB.105.094510} {\bibfield  {journal} {\bibinfo
   {journal} {Phys. Rev. B}\ }\textbf {\bibinfo {volume} {105}},\ \bibinfo
  {pages} {094510} (\bibinfo {year} {2022}{\natexlab{b}})}\BibitemShut
  {NoStop}%
\bibitem [{\citenamefont {Akhmerov}\ and\ \citenamefont
  {Beenakker}(2008)}]{PhysRevB.77.085423}%
  \BibitemOpen
  \bibfield  {author} {\bibinfo {author} {\bibfnamefont {A.~R.}\ \bibnamefont
  {Akhmerov}}\ and\ \bibinfo {author} {\bibfnamefont {C.~W.~J.}\ \bibnamefont
  {Beenakker}},\ }\bibfield  {title} {\bibinfo {title} {Boundary conditions for
  {Dirac} fermions on a terminated honeycomb lattice},\ }\href
  {https://doi.org/10.1103/PhysRevB.77.085423} {\bibfield  {journal} {\bibinfo
  {journal} {Phys. Rev. B}\ }\textbf {\bibinfo {volume} {77}},\ \bibinfo
  {pages} {085423} (\bibinfo {year} {2008})}\BibitemShut {NoStop}%
\bibitem [{\citenamefont {Basko}(2009)}]{PhysRevB.79.205428}%
  \BibitemOpen
  \bibfield  {author} {\bibinfo {author} {\bibfnamefont {D.~M.}\ \bibnamefont
  {Basko}},\ }\bibfield  {title} {\bibinfo {title} {Boundary problems for
  {Dirac} electrons and edge-assisted {Raman} scattering in graphene},\ }\href
  {https://doi.org/10.1103/PhysRevB.79.205428} {\bibfield  {journal} {\bibinfo
  {journal} {Phys. Rev. B}\ }\textbf {\bibinfo {volume} {79}},\ \bibinfo
  {pages} {205428} (\bibinfo {year} {2009})}\BibitemShut {NoStop}%
\bibitem [{\citenamefont {Davydova}\ \emph {et~al.}(2022)\citenamefont
  {Davydova}, \citenamefont {Prembabu},\ and\ \citenamefont {Fu}}]{dsp1}%
  \BibitemOpen
  \bibfield  {author} {\bibinfo {author} {\bibfnamefont {M.}~\bibnamefont
  {Davydova}}, \bibinfo {author} {\bibfnamefont {S.}~\bibnamefont {Prembabu}},\
  and\ \bibinfo {author} {\bibfnamefont {L.}~\bibnamefont {Fu}},\ }\bibfield
  {title} {\bibinfo {title} {Universal {Josephson} diode effect},\ }\href
  {https://doi.org/10.1126/sciadv.abo0309} {\bibfield  {journal} {\bibinfo
  {journal} {Science Advances}\ }\textbf {\bibinfo {volume} {8}},\ \bibinfo
  {pages} {eabo0309} (\bibinfo {year} {2022})}\BibitemShut {NoStop}%
\bibitem [{\citenamefont {Daido}\ \emph {et~al.}(2022)\citenamefont {Daido},
  \citenamefont {Ikeda},\ and\ \citenamefont
  {Yanase}}]{PhysRevLett.128.037001}%
  \BibitemOpen
  \bibfield  {author} {\bibinfo {author} {\bibfnamefont {A.}~\bibnamefont
  {Daido}}, \bibinfo {author} {\bibfnamefont {Y.}~\bibnamefont {Ikeda}},\ and\
  \bibinfo {author} {\bibfnamefont {Y.}~\bibnamefont {Yanase}},\ }\bibfield
  {title} {\bibinfo {title} {Intrinsic superconducting diode effect},\ }\href
  {https://doi.org/10.1103/PhysRevLett.128.037001} {\bibfield  {journal}
  {\bibinfo  {journal} {Phys. Rev. Lett.}\ }\textbf {\bibinfo {volume} {128}},\
  \bibinfo {pages} {037001} (\bibinfo {year} {2022})}\BibitemShut {NoStop}%
\end{thebibliography}
\end{document}

\title{Supplementary Material for ``Transverse Josephson diode effect in tilted Dirac systems''}

\author{W.~Zeng}
\email[E-mail: ]{zeng@ujs.edu.cn}
\affiliation{Department of physics, Jiangsu University, Zhenjiang 212013, China}

\maketitle

In this Supplemental Material, we present the calculation for constructing the Green's function using McMillan's formulation and provide the detailed derivations of the transverse current in the Josephson junction.

The Green's function can be constructed by the elementary scattering processes of the BdG Hamiltonian by the McMillan's formula \cite{PhysRev.175.559,PhysRevB.56.892,FURUSAKI1991299}. We set $\hbar=1$ throughout our calculation. The independent wave functions can be obtained by solving the scattering problem described by the secular equation $[E-H_{BdG}^\ell(\mathbf{r},\partial_\mathbf{r})]\psi(x)e^{ik_yy}=0$ with $k_y$ being the conversed transverse wave vector and $\ell=\pm$ being the valley index, yielding the scattering states 
\begin{align}
    &\psi_{1}(x)=A_{1}e^{+ik_1x}+a^\ell_{1}A_{4}e^{+ik_2x}+b_{1}^\ell A_{3}e^{-ik_1x},\label{eq:scatter1}\\
    &\psi_2(x)=A_2e^{-ik_2x}+a_2^\ell A_3e^{-ik_1x}+b_2^\ell A_4e^{+ik_2x},\\
    &\psi_3(x)=c_3^{\ell}A_1e^{-ik_1x}+d_3^{\ell}A_4e^{+ik_2x},\\
    &\psi_4(x)=c_4^{\ell}A_4e^{+ik_2x}+d_4^{\ell}A_3e^{-ik_1x},\label{eq:scatter2}
\end{align}
where $a^\ell,b^\ell,c^\ell,d^\ell$ are the scattering amplitudes. The basis spinors for the right-propagating electron-like quasiparticles (ELQ) and hole-like quasiparticles (HLQ) are given by
\begin{align}
    A_1=\begin{pmatrix}
        1\\\ell e^{i\ell\theta_1}\\\ell\gamma\\e^{i\ell\theta_1}\gamma
    \end{pmatrix},\quad
    A_2=\begin{pmatrix}
        -\ell e^{i\ell\theta_2}\gamma\\\gamma\\-e^{i\ell\theta_2}\\\ell
    \end{pmatrix},
\end{align}
respectively, whereas the basis spinors for the left-propagating ELQ and HLQ are given by
\begin{align}
    A_3=\begin{pmatrix}
        -\ell e^{i\ell\theta_1}\\1\\-e^{i\ell\theta_1}\gamma\\\ell\gamma
    \end{pmatrix},\quad 
    A_4=\begin{pmatrix}
        \gamma\\\ell e^{i\ell\theta_2}\gamma\\\ell\\e^{i\ell\theta_2}
    \end{pmatrix},
\end{align}
where $\gamma=\Delta/(E+\Omega)$ with $\Omega=\sqrt{E^2-\Delta^2}$ for $E>\Delta$ and $\Omega=i\sqrt{\Delta^2-E^2}$ for $E<\Delta$. The angle parameter is given by $\theta_{1,(2)}=\arctan(v_yk_y/v_xk_{1,(2)})$ with $k_{1,(2)}$ being the longitudinal wave vector for ELQ (HLQ), which is given by
\begin{align}
    k_{1,(2)}=v_x^{-1}\sqrt{(\mu_s-\ell v_t k_y+(-)\Omega)^2-(v_yk_y)^2}.
\end{align}

The conjugate Hamiltonian is given by $\tilde{H}_{BdG}^\ell(\mathbf{r},\partial_\mathbf{r})=[H_{BdG}^\ell(\mathbf{r},-\partial_\mathbf{r})]^T$, and the corresponding conjugate scattering states are governed by the secular equation $[E-\tilde{H}_{BdG}^\ell(\mathbf{r}^\prime,\partial_{\mathbf{r}^\prime})]\tilde{\psi}(x')e^{-ik_yy'}=0$, resulting in 
\begin{align}
    &\tilde\psi_{1}(x^\prime)=B_{1}e^{+ik_1x^\prime}+\tilde{a}^\ell_{1}B_{4}e^{+ik_2x^\prime}+\tilde{b}_{1}^\ell B_{3}e^{-ik_1x^\prime},\label{eq:scatter11}\\
    &\tilde\psi_2(x^\prime)=B_2e^{-ik_2x^\prime}+\tilde{a}_2^\ell B_3e^{-ik_1x^\prime}+\tilde{b}_2^\ell B_4e^{+ik_2x^\prime},\\
    &\tilde\psi_3(x^\prime)=\tilde{c}_3^{\ell}B_1e^{-ik_1x^\prime}+\tilde{d}_3^{\ell}B_4e^{+ik_2x^\prime},\\
    &\tilde\psi_4(x^\prime)=\tilde{c}_4^{\ell}B_4e^{+ik_2x^\prime}+\tilde{d}_4^{\ell}B_3e^{-ik_1x^\prime},\label{eq:scatter22}
\end{align}
where $\tilde{a}^\ell,\tilde{b}^\ell,\tilde{c}^\ell,\tilde{d}^\ell$ and $B_i$ ($i=1,2,3,4$) are the counterparts of $a^\ell,b^\ell,c^\ell,d^\ell$ and $A_i$, respectively. The scattering basis spinors ($B_i$) are given by
\begin{align}
    B_1=\begin{pmatrix}
        \ell e^{-i\ell\theta_1}\\-1\\\gamma e^{-i\ell\theta_1}\\\ell\gamma
    \end{pmatrix},\quad
    B_2=\begin{pmatrix}
        \gamma\\\ell\gamma e^{-i\ell\theta_2}\\\ell\\e^{-i\ell\theta_2}
    \end{pmatrix},\quad
    B_3=\begin{pmatrix}
        1\\\ell e^{-i\ell\theta_1}\\\ell\gamma\\\gamma e^{-i\ell\theta_1}
    \end{pmatrix},\quad
    B_4=\begin{pmatrix}
        \ell\gamma e^{-i\ell\theta_2}\\-\gamma\\e^{-i\ell\theta_2}\\-\ell
    \end{pmatrix}.
\end{align}

The retarded Green's function can be constructed by the scattering states obtained in Eqs.\ (\ref{eq:scatter1})-(\ref{eq:scatter2}) and (\ref{eq:scatter11})-(\ref{eq:scatter22}). We present the result for the left superconducting region ($x<0$). Due to the translational symmetry parallel to the junction, the retarded Green's function reads $G_\ell^R(x,x^\prime,y,y^\prime)=G_\ell^R(x,x^\prime)e^{ik_y(y-y^\prime)}$, where \cite{FURUSAKI1991299}
\begin{align}
    G_{\ell}^{R}(x,x')=\left\{
        \begin{array}{ll}
         \alpha_1\psi_1(x)\tilde\psi_3^T(x')+\alpha_2\psi_1(x)\tilde\psi_4^T(x')+\alpha_3\psi_2(x)\tilde\psi_3^T(x')+\alpha_4\psi_2(x)\tilde\psi_4^T(x'), & x>x', \\
          \beta_1\psi_3(x)\tilde\psi_1^T(x')+\beta_2\psi_4(x)\tilde\psi_1^T(x')+\beta_3\psi_3(x)\tilde\psi_2^T(x')+\beta_4\psi_4(x)\tilde\psi_2^T(x'), & x<x'.
        \end{array}
    \right.
\end{align}
The Green's function satisfies the following boundary condition obtained by integrating the BdG equation at $x=x'$, 
\begin{align}
     G_{\ell}^{R}(x+0^+,x)-G_{\ell}^{R}(x+0^-,x)=-iv_x^{-1}\ell\nu_z\sigma_x,
 \end{align} 
 which leads to the detailed balanced relations $a_1^\ell/\cos\theta_1=-\tilde{a}_2^\ell/\cos\theta_2$, $\tilde{a}_1^\ell/\cos\theta_1=-a_2^\ell/\cos\theta_2$, $b_{1,2}^\ell=\tilde{b}_{1,2}^\ell$, and the following identities
 \begin{gather}
 \alpha_1\tilde{c}^\ell_3+\alpha_2\tilde{d}^\ell_4=\beta_1c^\ell_3+\beta_2d^\ell_4=\frac{-i}{2v_x(1-\gamma^2)\cos\theta_1},\label{eq:coe1}\\
 \alpha_3\tilde{d}^\ell_3+\alpha_4\tilde{c}^\ell_4=\beta_3d^\ell_3+\beta_4c^\ell_4=\frac{i}{2v_x(1-\gamma^2)\cos\theta_2},\label{eq:coe2}\\
 \alpha_3\tilde{c}^\ell_3+\alpha_4\tilde{d}^\ell_4=\alpha_1\tilde{d}^\ell_3+\alpha_2\tilde{c}^\ell_4=\beta_1d^\ell_3+\beta_2c^\ell_4=\beta_3c^\ell_3+\beta_4d^\ell_4=0.\label{eq:coe3}
 \end{gather}
Consequently, with the help of Eqs.\ (\ref{eq:coe1})-(\ref{eq:coe3}), the solved retarded Green's function is given by
 \begin{align}
    G_{\ell}^{R}(x,x',E)=\frac{-i}{2v_x(1-\gamma^2)}\times\left\{
        \begin{array}{ll}
         \Xi_1^\ell/\cos\theta_1-\Xi_2^\ell/\cos\theta_2 & x>x', \\
          \tilde\Xi_1^\ell/\cos\theta_1-\tilde\Xi_2^\ell/\cos\theta_2, & x<x',
        \end{array}
    \right.\label{eq:fG}
\end{align}
where
\begin{align}
    &\Xi_1^\ell=A_1B_3^Te^{+ik_1(x-x')}+a^\ell_1A_4B_3^Te^{+ik_2x-ik_1x'}+b^\ell_1A_3B_3^Te^{-ik_1(x+x')},\\
    &\Xi_2^\ell=A_2B_4^Te^{-ik_2(x-x')}+a^\ell_2A_3B_4^Te^{-ik_1x+ik_2x'}+b^\ell_2A_4B_4^Te^{+ik_2(x+x')},\\
    &\tilde\Xi_1^\ell=A_3B_1^Te^{-ik_1(x-x')}+\tilde{a}^\ell_1A_3B_4^Te^{-ik_1x+ik_2x'}+\tilde{b}^\ell_1A_3B_3^Te^{-ik_1(x+x')},\\
    &\tilde\Xi_2^\ell=A_4B_2^Te^{+ik_2(x-x')}+\tilde{a}^\ell_2A_4B_3^Te^{+ik_2x-ik_1x'}+\tilde{b}^\ell_2A_4B_4^Te^{+ik_2(x+x')}.
\end{align}
Substituting $\gamma=\Delta/(E+\Omega)$ into Eq.\ (\ref{eq:fG}) gives rise to Eq.\ (4) in the main text.

With the help of the transverse current operator $\hat{j}_y=\sum_\ell \hat{j}_y^\ell$ with $\hat{j}_y^\ell=-e(v_y\psi_\ell^\dagger\sigma_y\psi_\ell+\ell v_t\psi_\ell^\dagger\psi_\ell)$, the transverse current $j_y^\ell$ can be expressed in terms of the Matsubara Green's function:
\begin{align}
    j_y^\ell=\langle\hat{j}_y^\ell\rangle=-\frac{e}{2\beta}\sum_{k_y,\omega_n}\Big(v_y\mathrm{tr}[\nu_0\sigma_yG_\ell(\mathbf{r},\mathbf{r},i\omega_n)]+\ell v_t\mathrm{tr}[\nu_0\sigma_0G_\ell(\mathbf{r},\mathbf{r},i\omega_n)]\Big).
\end{align}
The Matsubara Green's function for the BdG Hamiltonian $\mathcal{H}_{BdG}^\ell$ is defined by $G_\ell(\mathbf{r},\tau;\mathbf{r}',\tau')=-\langle\mathcal{T}\phi_\ell(\mathbf{r},\tau)\phi_\ell^\dagger(\mathbf{r}',\tau')\rangle$, where $\mathcal{T}$ is the time-ordering operator, $\phi_\ell=(\psi_\ell,\psi_{\bar{\ell}}^\dagger)^T$ is the Nambu spinor, and $\tau$ ($\tau'$) represents the imaginary time. 

For the positive frequency $\omega_n>0$, the Matsubara Green's function can be obtained by the analytic continuation $E+i 0^+\rightarrow i\omega_n$ from the retarded Green's function $G^R_\ell$, which has been already obtained by the scattering method. As a result, the expectation values of $\hat{j}_y^\ell$ are evaluated in terms of the Green's function with the summation over the Matsubara frequency and the wave number along the $y$ direction, which gives rise to the transverse current
\begin{align}
\langle \hat{j}_{y}^\ell \rangle_{\omega_n>0}=&\frac{1}{\beta}\sum_{k_y,\omega_n}J_{n}^{\ell,+},\\
J_n^{\ell,+}=&\frac{iev_y}{2}\Big(G_{\ell,21}(x,x,k_y,i\omega_n)-G_{\ell,12}(x,x,k_y,i\omega_n)+G_{\ell,43}(x,x,k_y,i\omega_n)-G_{\ell,34}(x,x,k_y,i\omega_n)\Big)\nonumber\\
    &-\frac{\ell ev_t}{2}\Big(G_{\ell,11}(x,x,k_y,i\omega_n)+G_{\ell,22}(x,x,k_y,i\omega_n)+G_{\ell,33}(x,x,k_y,i\omega_n)+G_{\ell,44}(x,x,k_y,i\omega_n)\Big).
\end{align}
With the help of Eq.\ (\ref{eq:fG}), one finds 
\begin{align}
    J_n^{\ell,+}=&\frac{\zeta_\ell e\Delta}{\Omega_n}\Big(\frac{a^\ell_1e^{-i\ell\theta_1}}{\cos\theta_1}+\frac{a^\ell_2e^{-i\ell\theta_2}}{\cos\theta_2}\Big)e^{2\Omega_nx/v_x}+\frac{e|\omega_n|}{\Omega_n}\Big(\frac{\zeta_1b_1e^{-2ik_1x}}{\cos\theta_1}-\frac{\zeta_2b_2e^{2ik_1x}}{\cos\theta_2}\Big)\nonumber\\
    &+\frac{ie|\omega_n|}{\Omega_n}\sum_{j=1,2}\Big(\zeta_j\tan\theta_j+\frac{\ell v_t}{2v_x}\cos\theta_j\Big),\quad (\omega_n>0),\label{eq:JN}
\end{align}
where $\Omega_n=\sqrt{\omega_n^2+\Delta^2}$, $\zeta_\ell=(4v_x)^{-1}[i\ell v_y(1-e^{i\ell(\theta_1+\theta_2)})+\ell v_t(e^{i\ell\theta_1}+e^{i\ell\theta_2})]$, and $\zeta_{1,2}=(2v_x)^{-1}(v_y+\ell v_t\sin\theta_{1,2})$. The transverse currents contributed by the Andreev and the normal reflection processes are shown in the first line of Eq.\ (\ref{eq:JN}), which are dependent on $a^\ell$ and $b^\ell$, respectively. The mean-field treatment of superconductivity requires the heavy-doping limit in the superconducting regions, \textit{i.e.}, $|\mu_s|\gg\{|\mu|,\Delta\}$ \cite{PhysRevLett.97.067007,RevModPhys.80.1337}. In this regime we have $\theta_1\simeq\theta_2\simeq\theta$ with $\theta=\arctan(v_yk_y/\mu_s)$, and $k_{1(2)}\simeq (\mu_s+(-)\Omega)/v_x$. The frequency component $J_n^{\ell,+}$ contributed by the Andreev processes show exponential decay into the bulk of the superconducting region on the scale of the coherence length $\xi_0=\hbar v_F/\Delta_0$. However, $J_n^{\ell,+}$ due to the normal reflection are modulated by a rapidly oscillating factor with the length scale $\sim (2\mu_s)^{-1}$, which is much small than $\xi_0$ in the heavy-doping regime. Consequently, the total current contributed by the normal reflection processes is negligible, resulting in
\begin{gather}
J_n^{\ell,+}=\frac{\zeta e\Delta}{2\Omega_n}\Big(a^\ell_1(\phi,i|\omega_n|)+a^\ell_2(\phi,i|\omega_n|)\Big)e^{2\Omega_nx/v_x}+\frac{i\zeta e|\omega_n|}{\Omega_n},\\
\zeta=\frac{v_y\tan\theta+\ell v_t\sec\theta}{v_x}.\label{eq:coef}
\end{gather}
For the negative frequency $\omega_n<0$, the Matsubara Green's function can be obtained by analytic continuation from the advanced Green's function, \textit{i.e.}, $G^A(E-i0^+)\rightarrow G(-i|\omega_n|)$, where the advanced Green's function $G^A$ can be obtained by the similar procedure as we obtain $G^R$. The transverse current contributed by the negative frequency is given by
\begin{align}
\langle \hat{j}_{y}^\ell \rangle_{\omega_n<0}&=\frac{1}{\beta}\sum_{k_y,\omega_n}J_{n}^{\ell,-},\\
J_n^{\ell,-}&=\frac{\zeta e\Delta}{2\Omega_n}\Big(\bar{a}^\ell_1(\phi,-i|\omega_n|)+\bar{a}^\ell_2(\phi,-i|\omega_n|)\Big)e^{2\Omega_nx/v_x}-\frac{i\zeta e|\omega_n|}{\Omega_n}.
\end{align}
For a finite $|\omega_n|$, one finds that the imaginary parts of $J_n^{\ell,+}$ and $J_n^{\ell,-}$ are equal in magnitude but opposite in sign, whereas their real parts are the same. Then, the resulting transverse current density is given by
\begin{align}
    j_{y}^\ell=&\langle \hat{j}_{y}^\ell \rangle_{\omega_n>0}+\langle \hat{j}_{y}^\ell \rangle_{\omega_n<0}=\frac{1}{\beta}\sum_{k_y,\omega_n>0}\Big(J_{n}^{\ell,+}+J_{n}^{\ell,-}\Big)\nonumber\\
    =&\frac{1}{\beta}\sum_{k_y,\omega_n>0}\frac{\zeta e\Delta}{\Omega_n}\Big(\mathfrak{Re} \bm{[}a^\ell_1(\phi,i\omega_n)\bm{]}+\mathfrak{Re} \bm{[}a^\ell_2(\phi,i\omega_n)\bm{]}\Big)e^{2\Omega_nx/v_x}.\label{eq:tty}
\end{align}

\begin{figure}[tp]
\begin{center}
\includegraphics[clip = true, width =0.5\columnwidth]{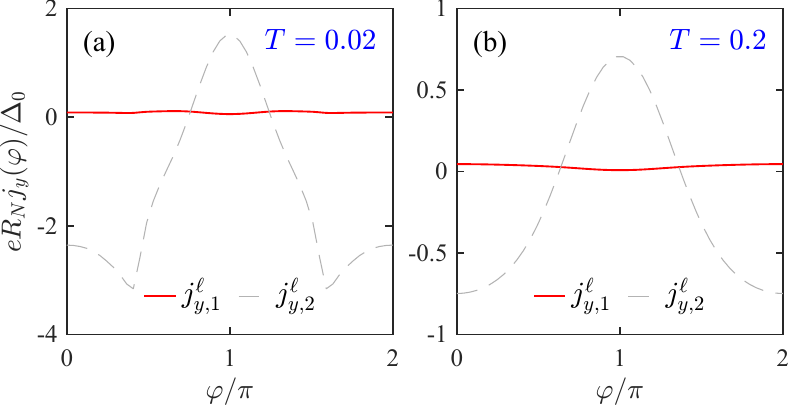}
\end{center}
\caption{Current-phase relation of $j_y^\ell(\varphi)$ at $\kappa=1.2$. (a) $T=0.02$ (in units of $T_c$). (b) $T=0.2$.}
\label{fig:s1}
\end{figure}

We note that the dimensionless coefficient in Eq.\ (\ref{eq:coef}) can be rewritten as $\zeta=\zeta_1+\zeta_2$ with $\zeta_1=(v_y/v_x)\tan\theta$ and $\zeta_2=(\ell v_t/v_x)\sec\theta$, which are related to the direction-dependent velocity $v_y$ and the tilt velocity $v_t$, respectively. As a result, the transverse current in Eq.\ (\ref{eq:tty}) can be decomposed into two parts, \textit{i.e.}, $j_y^\ell=j_{y,1}^\ell+j_{y,2}^\ell$, where $j_{y,1,(2)}^\ell=k_BT\sum_{k_y,\omega_n}\zeta_{1,(2)}e\Delta/\Omega_n\mathfrak{Re}(a_1^\ell+a_2^\ell)e^{2\Omega_nx/v_x}$. In the heavy-doping regime, we have 
\begin{align}
    \frac{\zeta_1}{\zeta_2}=\frac{v_y}{v_t}\sin\theta\sim (v_y/v_t)\left|\frac{\mu}{\mu_s}\right|\ll1,
\end{align}
which results in $j_{y,1}^\ell\ll j_{y,2}^\ell$. The current-phase relation of $j_{y}^\ell$ is shown in Figure.\ \ref{fig:s1}. It is shown that $j_{y,1}^\ell$ is significantly smaller than $j_{y,2}^\ell$. We note that $j_{y,1}^\ell$ is related to the skew-tunneling and is absent when the scattering is symmetric, \text{i.e.}, $a_{1,2}^\ell(k_y)=a_{1,2}^\ell(-k_y)$. Whereas $j_{y,2}^\ell$ is attributed to the tilt-induced nonzero transverse momentum acquired by the quasiparticles. Consequently, the predicted transverse Josephson Hall current is dominated by the passage of the Cooper pair across the junction carrying a finite tilt-induced transverse momentum.